\documentclass[%
 reprint,
 amsmath,amssymb,
 aps,
pra,
]{revtex4-2}
\usepackage[toc,page]{appendix}

\usepackage[utf8]{inputenc}
\usepackage{bm}
\usepackage{multirow,amssymb,amsbsy,amsmath}
\usepackage{graphicx}% Include figure files
\usepackage{subfigure}
\usepackage{mathrsfs}
\usepackage{verbatim}
\usepackage{amssymb}
\usepackage{amsfonts}
\usepackage{bbm}
\usepackage{amsmath}
\usepackage{lmodern}
\usepackage{appendix}
\usepackage{soul}
\usepackage[colorlinks,citecolor=blue,urlcolor=blue, linkcolor=blue]{hyperref}
\usepackage{ulem}
\usepackage{xcolor}
\usepackage{sidecap}
\usepackage{dcolumn}% Align table columns on decimal point
\usepackage{bm}% bold math
\newcommand{\tr}{{\rm tr}}

\newcommand{\m }{\mathcal}

\newcommand{\bra}[1]{\langle#1|}
\newcommand{\ket}[1]{|#1\rangle}

\newcommand{\mean}[1]{\langle #1 \rangle}

\newcommand{\proj}[1]{\ket{#1}\!\bra{#1}}

\newcommand{\Stephan}[1]{{\color{black} #1}}
\newcommand{\Ali}[1]{{\color{black} #1}}
\begin{document}

%\preprint{APS/123-QED}

\title{Covariant correlation-disturbance and its experimental realization \\with spin-1/2  particles}

\author{Ali Asadian$^1$, Florian Gams$^2$, and Stephan Sponar$^2$}

\affiliation{$^1$Department of Physics, Institute for Advanced Studies in Basic Sciences (IASBS), Gava Zang, Zanjan 45137-66731, Iran\\
$^2$Atominstitut, TU Wien, Stadionallee 2, 1020 Vienna, Austria 
}

\begin{abstract}
%We formulate a precise tradeoff relation between correlation and disturbance in a sequential quantum measurement setup.

We formulate a precise tradeoff relation between correlation and disturbance for \Stephan{sequential $n$-outcome quantum measurements in Hilbert spaces of arbitrary dimension.}
This relation
%, examined for qubit case, 
highlights key symmetry properties useful for robust estimation and characterization of the measurement parameters against unitary noise, or in
scenarios where shared reference frames are unavailable. In addition, we report on the experimental implementation of the proposal \Stephan{for the qubit case, more precisely} in a neutron optical experiment, which is particularly valuable for calibrating and optimizing measurement devices, as confirmed by the theoretical results. \Stephan{Finally, we exploit the optimal tradeoff relation for direct estimation of the characteristic noise of single-photon detectors, dark counts and the finite detection efficiency.}

\end{abstract}

\maketitle

%\tableofcontentshttps://www.overleaf.com/project/6420e559e3a63b9e2bee4b89

% Q incomp Uncertainty & recources
%%%%%%%%%%%%%%%%%%%%%%%%%%%%%%%%%%%%%%%%%%%%%%%%%%%%%
{\it Introduction.---} 
%General%%%%%%%%%%%%%%%%%%%%%%%%%%%%%%%
Incompatibility between measurements is a fundamental aspect of quantum mechanics, leading to distinctive features such as nontrivial uncertainty \cite{Heisenberg27, RMP2} and complementarity \cite{Bohr28} relations. At its core, quantum violations of various classical models--commonly referred to as hidden variable models \cite{Bell64,Bell66,Kochen67}--necessitate this measurement incompatibility, which can be verified solely through observed statistics without relying on specific assumptions about the devices \cite{Wolf}.  Identifying scenarios that exhibit the characteristic features of measurement incompatibility and disturbance is now a central objective for both fundamental research and practical applications, as it has been recognized as a key resource underlying many quantum communication tasks that outperform classical methods \cite{RMP1, Arute2019, Sen}. Understanding and controlling measurement disturbance has far-reaching implications for developing more precise and reliable quantum technologies \cite{Kwiat, Oszmaniec,Elben2023}.

%The variety of ways of analyzing quantum incompatibility involves realizing correlation measurements between spatially separated systems, as is the case in Bell nonlocality test, or temporal sequence of measurements performed on a single system. The latter is used  for realizing the quantum violation of macroscopic realism and non-contextuality.

%Significance & appication%%%%%%%%%%%%%%%%%
%\aa{This parag should be removed? An interesting variant of realizing quantum uncertainty is illustrated through sequential measurements. [quantum communication task as...] ...Uncertainty relation, error-disturbance is one of crucial way of ...Recently the resource theory of ...}

%Highihgt  the approaches (Error-Distur)%%%%%
One of the most fundamental approaches to quantum complementarity and uncertainty is investigated via a temporal sequence of measurements performed on an initially prepared single system. Many interesting quantum predictions, such as quantum contextuality and the violation of macroscopic realism,  have been demonstrated in this scenario~\cite{LGI85,Emary_2014,Knee2016,Sponar2024, LGreview}.
%\aa{[mention key words: measurement disturbance, NSIT]}. 
The interplay between information gain and disturbance has been formulated in various tradeoff relation inequalities \cite{Asher1996}.  The first information theoretic - or entropic - uncertainty relation was formulated by Hirschman \cite{Hirschman57} in 1957 for the position and momentum observables, and later improved in Refs \cite{Beckner75, Birula75}. The extension to non-degenerate observables on a finite-dimensional Hilbert space was then proposed by Deutsch in 1983 \cite{Deutsch83} and later improved by Maassen and Uffink \cite{Maassen88} yielding the well-known entropic uncertainty relation $H( A)+H( B)\geq -2\,{\rm log_2}\,c$, where $H$ denotes the Shannon entropy and $c$ is the maximal overlap between the eigenvectors $\ket {a_i}$ and $\ket{ b_j}$ of the observables $A$ and $B$. Entropic uncertainty has useful applications in entanglement witnessing~\cite{Berta2010}, complementarity~\cite{Coles2014}, and in quantum information theory~\cite{NielsenChuang}. More general procedures aiming to quantify error and disturbance in successive (or simultaneous) measurements are Ozawa's operator-based measures between target observables and measurements~\cite{Ozawa03,Ozawa04} or distances of the associated probability distributions~\cite{Werner04}. Interest has risen in alternative information-theoretic measures, introduced recently by Buscemi ~\cite{Buscemi14}, and in several subsequent alternative approaches~\cite{Coles15,Baek16,Schwonnek16,Barchielli18, Saberian}.

% Our approach & advantage%%%%%%%%%%%%%%%%%
This work proposes a precise tradeoff relation between correlation and disturbance within a straightforward, Ramsey-like successive measurement setup. Our scheme effectively captures the inherent incompatibility between two general measurements through a covariant tradeoff complementarity inequality, where the characteristic measurement values remain invariant under coordinate choices or unitary transformations of the basis.
In contrast to other approaches \cite{Buscemi14, Busch14}, our tradeoff relation relies on definitions that are both functionally simple and experimentally practical. This provides a clear and efficient method for robust experimental estimation of measurement devices, enabling tasks such as self-calibration and assessing quantum measurement performance. Additionally, our framework supports various communication protocols, including quantum random access codes and quantum key distribution (QKD), making it suitable for (semi) device-independent quantum communication scenarios, as explored in numerous studies. \cite{Mohan2019, Miklin, Anwer, Roy2023, PRXQ, Liang,Elben2023,Gherardini24}.

%%%%%%%%%%%%%%%%%%%%%%%%%%%%%%%%%%%%%%%
{\it Setting.---} 
%General%%%%%%%%%%%%%%%%%%%%%%%%%%%%%%%
Our \Stephan{proposed scheme} consists of three parts:  a state preparation $\rho$ followed by two successive observable measurements, denoted by $\m M_a=\sum_\alpha \alpha E_{\alpha|a}$ and $\m M_b=\sum_\beta \beta E_{\beta|b}$, respectively. Formally, the general measurements are described by a set of positive operator-valued measures (POVMs) where $E_{\alpha|a}\geq 0$ and $\sum_a E_{\alpha|a}=\mathbbm 1$  (see, Fig. 1). As is clear from the figure the measurement device $\m M_a$  performs a general measurement operation that not only produces outcomes (classical output) but also realizes the corresponding post-measurement state, according to the transformation rule, described by $\rho\mapsto \m I_{\alpha|a}(\rho)=K_{\alpha|a}\rho K_{\alpha|a}^\dag $ with outcome probability $p_{\alpha|a}=\tr [\m I_{\alpha|a}(\rho)]$ and $E_{\alpha|a}=K_{\alpha|a}^\dag K_{\alpha|a}$. Summing over the Kraus operators,  i.e., $\m I_a(\rho)=\sum_\alpha  \m I_{\alpha|a}(\rho)$, acts as a quantum channel corresponding to the case when the measurement is performed but the outcome is ignored or unregistered. This situation, however, affects the outcome statistics of the subsequent measurement. Therefore,
quantum instruments facilitate the analysis of measurement disturbance and incompatibility in sequential measurement setups. The important fact is that the Kraus operators ($K_{\alpha|a}$), generally given by $K_{\alpha|a}=U_{a|\alpha}E_{\alpha|a}^{\frac{1}{2}}$, are not unique.  We are mainly interested in a particular choice of the instrument associated with the observable measurement effect only and not the unitary afterward, that is $K_{\alpha|a}=E_{\alpha|a}^{\frac{1}{2}}$ so-called L\"uder instrument which merely captures the intrinsic effect of the observable's POVM in the state update rule. Therefore, this special case of instrument is more suitable to study the incompatibility and joint measurability of the observables in sequential implementations \cite{Teiko}.

%%%%%%%%%%%%%%%%
\begin{figure}[t!]
\includegraphics[width=9cm]{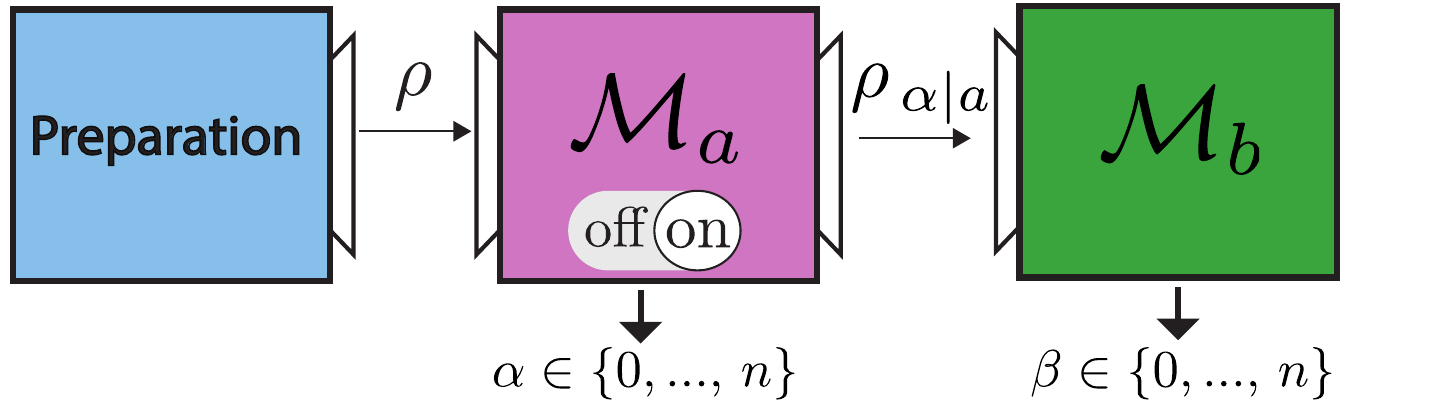}	
\caption{Ramsey-like interferometric sequence. Schematics of the setup involving three parts, state preparation, and successive \Stephan{$n$-outcome} measurements $\mathcal M_a$ (probe/reference), and measurement $\mathcal M_b$ (target). For measuring the measurement disturbance $\m M_a$ is turned on (performed) and off (unperformed). }\label{fig1} 
\end{figure}
%%%%%%%%%%%%%%%%%

%%%%%%%%%%%%%%%%%%
{\it Correlation-disturbance tradeoff high-dimensional case}.---
%%%%%%%%%%%%%%%%%%
\Stephan{First we want to emphasize, that the following derivations are general and apply to $n$-outcome measurements in a Hilbert space of arbitrary dimension.} Central to our formulation/construction is the Born rule for evaluating the joint probabilities of the two measurement outcomes: 
\begin{equation}
\label{joint}
p(\alpha,\beta)={\rm tr}\Big[ \mathcal I_{\alpha|a}(\rho) E_{\beta| b}\Big],
\end{equation}
which encapsulates the essential information required to analyze the interplay between the two measurements. First, the above form fulfills, the so-called arrow of time, i.e., $p(\alpha|a)=\sum_\beta p(\alpha,\beta|a, b)= \sum_\beta p(\alpha,\beta|a, b')$ for all $b\neq b'$, namely, the future measurement will not affect the past measurement. 

More characteristically, we have:
\begin{equation}
  \tilde p(\beta|b)= \tr[\rho \m I^*_a(E_{\beta|b})]=\sum_{\alpha} p(\alpha,\beta),
\end{equation}
which is equivalent to the scenario where the $\m M_a$ measurement is performed (or turned on) but its outcome is not recorded. Note that this is different from the case in which measurement $\m M_a$ is unperformed (turned off), $p(b|y)=\tr[\rho E_{b|y}]$. This means that the difference in the probabilities of obtaining outcome \( \beta \) when it is preceded by the (unregistered) \( \mathcal{M}_a \) measurement, compared to when \( \mathcal{M}_b \) is measured alone, reflects the measurement disturbance \cite{Roope, Emary, RMP1, Hamed}.
Therefore, a standard distance measure quantifying the disturbance may given by
\begin{align}
   \Ali{ \m D=\sqrt{\frac{n}{n-1}}\parallel \vec p(\beta|b)- \vec{\tilde p}(\beta|b)\parallel_2,} %=&|\sum_\beta \beta p(\beta|b)|-|\sum_\beta \beta \tilde p(\beta|b)|\leq \m D
\end{align}
where $n$ is the number of different outcomes.
Note that the value of $\m D$  depends on the initial state. 
The maximum value corresponds to the spectral norm, $\m D\leq {\rm sup}_\rho |\m D_{\rho}|\leq 1$, realized for the optimal state maximizing the disturbance detection. The effect of the probe measurement on the target measurement is described by the L\"uders channel modifying the observable $b$ (target), i.e., $\m I^*_a(\m M_b)=\sum_\alpha E_{\alpha|a}^{\frac{1}{2}}\m M_b E_{\alpha|a}^{\frac{1}{2}} $. Here $\m I^*_a$ is the dual of a quantum channel preserving the identity operator, $\m I^*_x(\mathbbm 1)=\mathbbm 1$. \Ali{The maximum disturbance 1 is obtained for $\m M_a$ and $\m M_b$ being complimentary, two different mutually unbiased bases, measurements. }

%This implies that the effect is completely characterized by the Measurement's Bloch vectors which simplifies the characterization . Its effect can be investigated on the \aa{In the context of Leggett-Garg Macrorealism $d_y(\rho)=0$ is called NSIT condition. Therefore, no-signaling in time is violated by the above form. It is also considered as quantum witness in []. It appears to be a natural way of measuring measurement disturbance in other works [ ]}  

The other measure of interest, defined in terms of Eq. \eqref{joint} quantifies the correlation between the successive outcomes,
\begin{equation}
  \Ali{   \m C=\frac{n}{n-1}\Big( p(\alpha=\beta)- \frac{1}{n}\Big)}
\end{equation}
%function:
%\begin{align}\label{eq:Corr}
%    \m C=\mean{\m M_a \m M_b}_{\rm seq}=\sum_{\alpha,\beta} \alpha\beta \,p(\alpha,\beta|a,b)=\sum_\alpha \alpha \mean{\m I^*_{\alpha|a}(\m M_b)}_\rho
%\end{align}
whose absolute value indicates the predictability of the target measurement given the result of the probe(reference) measurement.  

One naturally expects that the measurement disturbance suppresses the predictability and, therefore, expects a mutually complementary role between the defined quantities.
In the following, we report fundamental statistical constraints between the outcomes of sequential measurements expressed in the form of a tradeoff inequality between correlation and disturbance:
\begin{equation}\label{eq:main}
    \m C^2+ \m D^2\leq 1.
\end{equation}
This inequality, we call $\m C \m D$-tradeoff, constituting the starting point of our main result, represents a nontrivial mutual relation: a large correlation implies a small disturbance and vice versa. The sketch of the proof is presented in Supplemental Material \cite{Supp} Sec. I. The mathematical form of \eqref{eq:main} appears in a different context known as wave-particle duality relation which has been proposed to illustrate the complementarity between interference visibility and path distinguishability as a quantum feature independent of Heisenberg uncertainty or noncommutativity \cite{Englert96}. In our scheme, however, uncertainty is captured via the measurement disturbance. \Ali{ Although the definitions extend to multiple-outcome measurements, in many cases two-outcome measurements sufficiently capture the essential features. Therefore, we focus on dichotomic measurements, where $\alpha, \beta = \pm 1$. In this case, the expressions for correlation and disturbance are simplified to
$\m C=2p(\alpha=\beta)-1$ and $\m D=2|p(+|b)-\tilde p(+|b)|$.}

To gain a clear intuition about various specifications and applications of \eqref{eq:main}, we examine the qubit example and present a precise $\m{CD}$-tradeoff holding covariant symmetry proving to be highly useful for the experimental characterization of measurement parameters.

%%%%%%%%%%%%%%%%%%%
{\it Qubit case and the covariant symmetry.---}
%%%%%%%%%%%%%%%%%%
The initial state of a qubit in Bloch representation is  $\rho=\frac{1}{2}(\mathbbm 1+\vec r\cdot \vec\sigma)$ where $\vec r$ is called Bloch vector. The most general form of the two-outcome POVMs for target measurement is $E_{\pm|b}=((1\pm b_0)\mathbbm 1\pm \vec b\cdot \sigma )/2$ giving $\m M_b=E_{+|b}-E_{-|b}=b_0\mathbbm 1+\vec b\cdot \vec\sigma $ in which $b_0$ and $|\vec b|$ are interpreted as the measurement \textit{bias} and \textit{sharpness} (strength), respectively, as characteristic parameters describing the measurement apparatus. The positivity condition imposes $|b_0|+|\vec b|\leq 1$. Given the Bloch representation of the POVM elements and the L\"uders channel, the disturbance takes the form $\m D=\mean{\vec b\cdot \vec\sigma}_\rho-\mean{\vec {\tilde b}\cdot \vec\sigma)}_\rho=(\vec b-\vec{\tilde b})\cdot \vec r$ where  $\vec {\tilde b}\cdot \vec\sigma=\m I^*_a(\vec b\cdot \vec\sigma)$. 
As expected, the amount of disturbance depends on the noncommutativity between the corresponding observables, $\m D=-s |\vec b| \mean{\big[ \hat a\cdot \vec\sigma, [\hat a\cdot \vec\sigma, \hat b\cdot \vec\sigma]\big]}_\rho=-s  |\vec b| \hat a\times(\hat a \times \hat b)\cdot \vec r$. Here, the coefficient $s=1-(u_+ +u_-)/2$ where, $u_\alpha=\sqrt{(1+\alpha a_0)^2-|\vec a|^2}$ accounts for the outcome-dependent \textit{unsharpness} of $\m M_a$ measurement.
The above form suggests that the optimal input state maximizing $\m D$ is obtained by a unit Bloch vector $\vec r_{\rm opt}=(\vec b- \vec {\tilde b})/|\vec b- \vec {\tilde b}|=-\hat a\times(\hat a \times \hat b)$ which lies in the plane spanned by the measurements $\m M_a$ and $\m M_b$ and is perpendicular to the measurement direction of $\m M_a$, as depicted in Fig.\,\ref{fig:CorrVsDist}\,(b), resulting to $\mean{\m M_a}=a_0$. In this case, $\m D=s  |\vec b|\sin\theta$ where $\theta=\arccos (\hat a\cdot \hat b)$ represents the angle between the Bloch vectors of the respective measurements.  Correlation for the optimal state is expressed as $\m C=a_0b_0+|\vec a||\vec b|\cos\theta+\delta|\vec b|\sin\theta$. As an example, consider $\m M_x$ being sharp and the second measurement arbitrary then Eq.\eqref{ellipse} reduces to an optimal tradeoff,  $\m C^2+\m D^2=|\vec b|^2$. The practical advantage of this finding becomes more significant in high-dimensional measurement in which the readout noise or detector precision as represented by the generalized Bloch length is estimated without resorting to highly demanding process tomography(see \cite{Supp} for details in $d$-dimension).
Therefore, one can directly measure the strength of the second measurement by estimating the correlation and disturbance. Note that the form of the relation is invariant under the choice of the measurement settings or the observer's frame of reference, hence called covariant correlation-disturbance. 

Consequently, the entire result for the most general pair of measurements can be summarized into the following parametric equations governing the $\m{CD}$-tradeoff:
\begin{equation}
\label{ellipse}
 \left[\begin{array}{c} \m C \\ \m D \end{array}\right]=\left[\begin{array}{c} a_0b_0 \\0\end{array}\right]+\left[\begin{array}{cc}|\vec a| & \delta \\ 0 & s \end{array}\right]\left[\begin{array}{c} |\vec b|\cos \theta \\ |\vec b|\sin \theta \end{array}\right]
\end{equation}
where $\delta=(u_+ -u_-)/2$ is non-zero for bias of $\m M_a$, introducing shear to the original circle. The bias of the target measurement is inferred from the displacement $a_0 b_0$ along the $x$-axis. Thus,
the values of $\m C$ and $\m D$ follows an ellipse equation
encapsulating all essential information about the intrinsic properties of the measurement devices in a generic form, regardless of the unitary transformation of the observables, which means that measurement settings (e.g. measurement direction $\theta$) do not have to be known in principle. Therefore our scheme offers an experimentally robust estimation and characterization of the measurement parameters such as the measurement strength even with limited control over the measurements, or in situations involving uncharacterized local unitary noise. 
\Stephan{Note that in comparison to the scheme proposed by Buscemi \cite{Buscemi14}, that is also based on correlations (but between initial states and measurement outcomes), no prior knowledge about the observables - or more precisely - about its eigen states is required here. This paves the way for self-calibration of quantum measurement devices in (semi) device-independent quantum communication applications.}

%%%%%%%%%%%%%%%%
\begin{figure}[!b]
\includegraphics[width=8.5 cm]{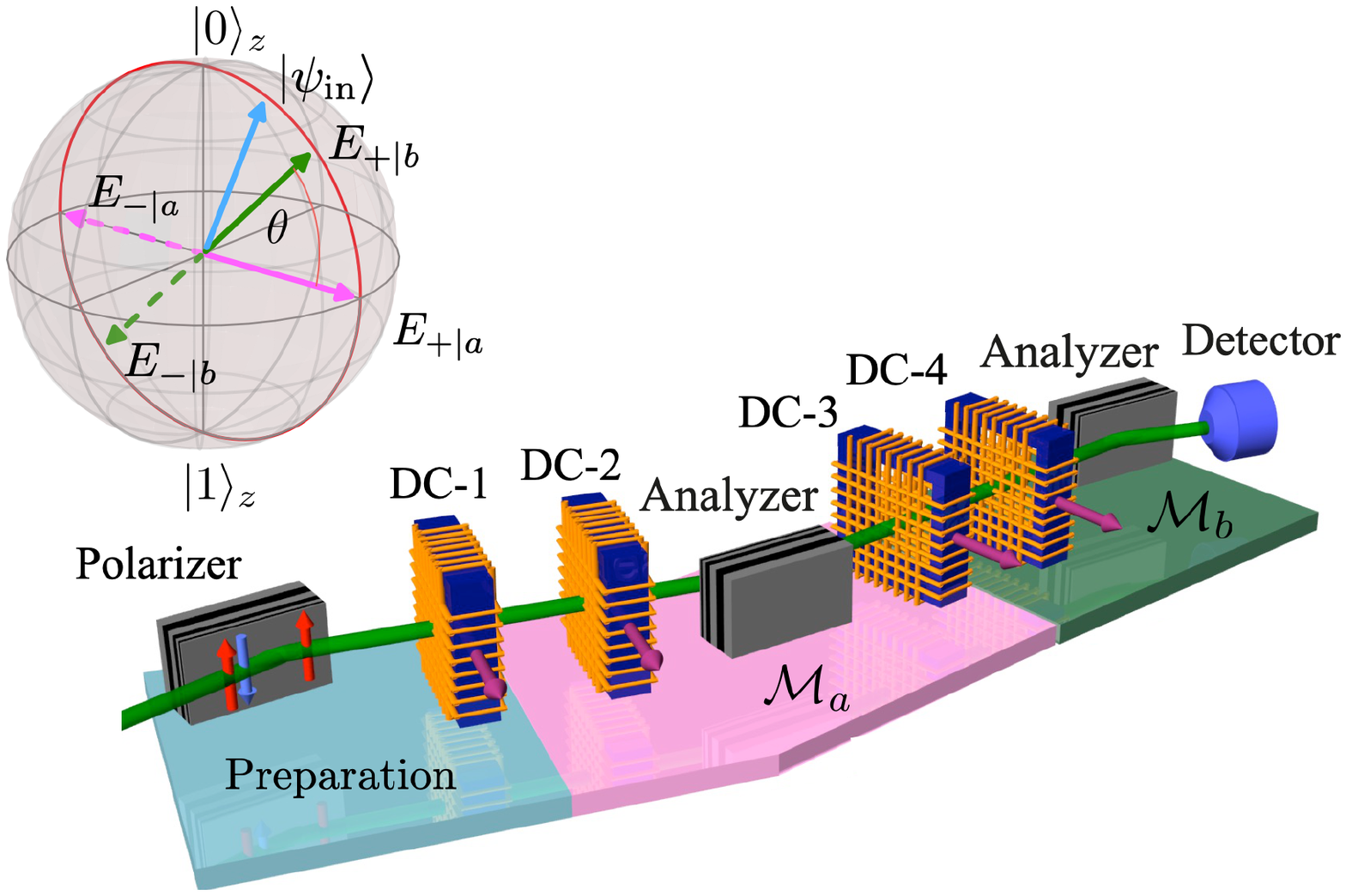}	
\caption{Neutron optical setup for observing $\m{CD}$-tradeoff using supermirrors (polarizers, analyzers) as projection operators and exploiting Larmor recession induced by the magnetic fields and spin turner coils DC-1 to DC-4, any required manipulation of the neutron spin can be performed. Bloch sphere indicates initial \textcolor{blue}{(pure)} state $\ket{\psi_{\rm{in}}}$, and measurement direction of the first $\m M_{a}$ and second measurement $\m M_{b}$, respectively.}\label{fig:setup} 
\end{figure}
%%%%%%%%%%%%%%%%%

%\aa{question: is $\vartheta=\arccos(\m C_{xy}/ \m S)$ or $\vartheta=\arctan(\m D_{y\tilde y}/\m C_{xy})$ is always corresponds to $\arccos (\vec x\cdot \vec y/|\vec x| |\vec y|)$}

{\it Experimental results.---} We experimentally demonstrate various instances of the above relationship. The data offers significant validation and verification of the measurement setup's performance by leveraging the covariance symmetry of the $\m{CD}$-tradeoff.
%The experiment was performed at the 250 kW TRIGA Mark-II research reactor at TU Wien, Austria. 
The configuration of the experimental setup is depicted in Fig.\,\ref{fig:setup} (see Supplementary Material Sec. III\,A for details). \Ali{In all our experimental configurations we examine the effects of the first measurement $\m M_{a}$ on the subsequent measurement $\m M_{b}$ for the optimal  \textcolor{blue}{(pure)} input state $\ket{\psi_{\rm opt}}$, in terms of disturbance $\m D$ and correlations $\m C$. }

For a better understanding of the implementation of a general measurement process in our setup, it is useful to rewrite the POVM in the following form 
\begin{equation}\label{eq:POVMbias}
E_{\pm\vert b}(\theta,\gamma,b_0)=\gamma\Pi_\pm(\theta)+(1-\gamma)N_\pm(b_0)
\end{equation}
as a convex combination of projective measurement $\Pi_\pm(\theta)=(\mathbbm 1\pm\hat b(\theta)\cdot\vec\sigma)/2$, with unit vector $\hat b(\theta)$, and biased dummy (noisy) measurement $N_\pm(b_0)=\frac{1}{2}(1\pm\frac{b_0}{1-\gamma})\mathbbm 1$, which is fully characterized by three parameters, rotation angle $\theta$ about a specific axes, bias $b_0$, and the randomization degree $\gamma=\vert \vec b\vert$ (see, Sec. III. D in \cite{Supp} for the experimental implementation of POVMs).

First, we study the case where the probe (first) measurement is projective and given by $\m M_{a}=\sigma_x$. In contrast, the target (second) measurement is represented by the unbiased POVM $\m M_{b}(\gamma,\theta)=E_{+\vert b}(\gamma,\theta)-E_{-\vert b}(\gamma,\theta)$, with POVM elements $ \Pi_{\pm\vert b}(\gamma,\theta)$ from Eq. (\ref{eq:POVMbias}). The resulting tight trade off relation (Eq. (\ref{ellipse})) $\m C^2+\m D^2=\vert \vec b\vert^2$ can be seen in Fig.\,\ref{fig:CorrVsDist} \Stephan{(the special case $\m C^2+\m D^2=1$ with $\gamma=1$, where both measurements are {\it{projective}}, is plotted in blue)}. This particular configurations allows for a direct determination of the measurement strength $\gamma$ and in turn for the Bloch length $\vert\vec b\vert$. This self-calibration features results in the following experimentally determined interaction strengths $\gamma$:    $\gamma_1=0.731(4)$, $\gamma_2=0.485(3)$, and $\gamma_2=0.233(3)$, which is plotted in Fig.\,\ref{fig:CorrVsDist} in magenta, violet and red, respectively. In a configuration where the POVM is the last measurement the bias $b_0$ has no effect. 
%%%%%%%%%%%%%%%%
\begin{figure}[h!]
\includegraphics[width=8.5 cm]{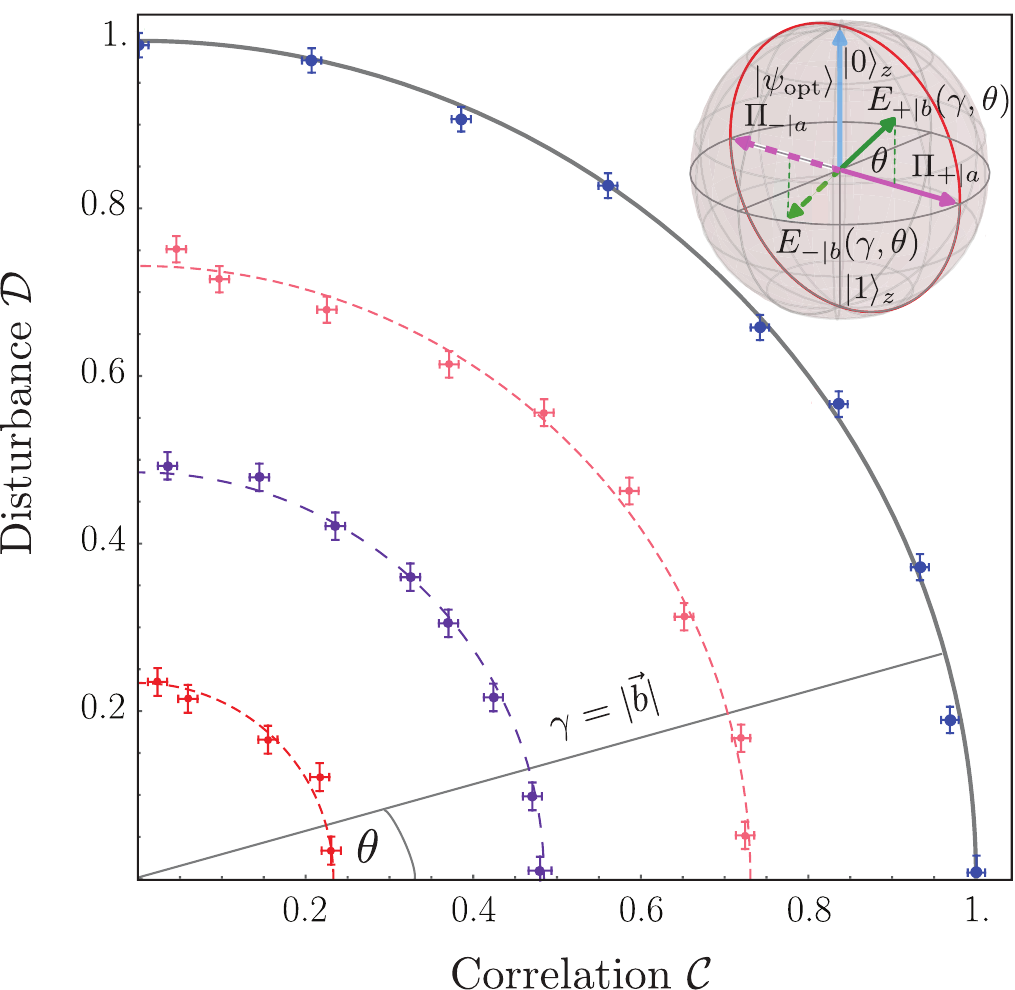}	
\caption{Disturbance $\m D$ versus correlation $\m C$ first (fixed) projective $\m M_{a}=\sigma_x$ and second (rotated) unbiased POVM $\m M_{b}(\gamma,\theta)=E_{+\vert b}(\gamma,\theta)-E_{-\vert b}(\gamma,\theta)$. The measurement strength $\gamma$ equals the length of Bloch vector $\vec b$ of the target measurement $\m M_{b}$. %The angle from the horizontal line corresponds to $\vartheta=\arccos (\vec a\cdot \vec b/|\vec a| |\vec b|)$ and the angular distance $D_\vartheta=2\vartheta/\pi$ separates the different points. The angle change by rotation about $\hat n$ with angle $\theta$ and $\vartheta_f=\theta+\vartheta_i$.
}\label{fig:CorrVsDist} 
\end{figure}
%%%%%%%%%%%%%%%%%

\begin{figure}[!t]
\includegraphics[width=8.25 cm]{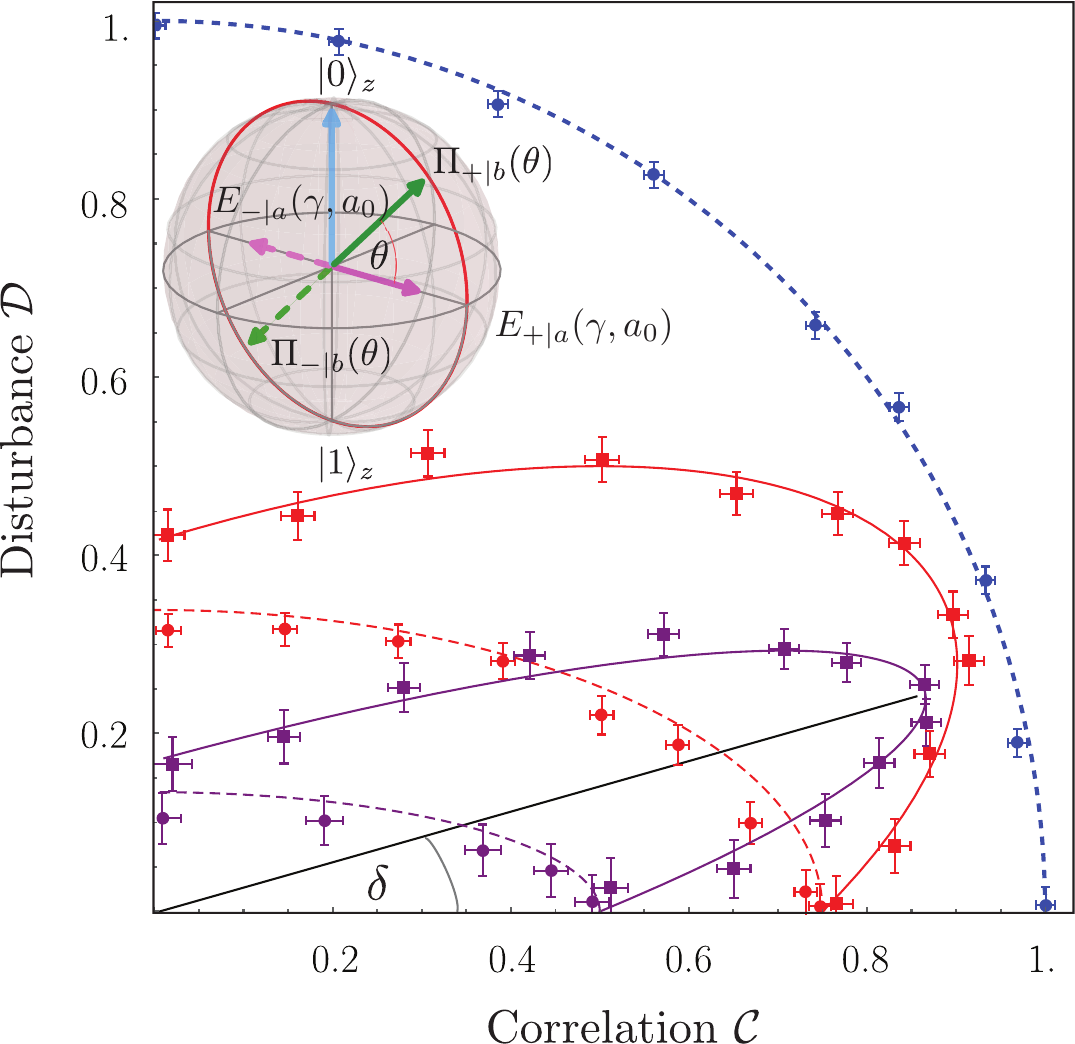}	
\caption{Disturbance $\m D$ versus correlation $\m C$ with first unbiased and biased POVM $\m M_{a}(\gamma,\,a_0)=E_{+\vert a}(\gamma,a_0)-E_{-\vert a}(\gamma,a_0)$ and second (varying) projective $\m M_{b}(\theta)=\cos\theta\sigma_x+\sin\theta\sigma_z$. \Stephan{The length of Bloch vector $\vec a$ of the first measurement accounts for the maximal possible correlation $\m C_{\rm{max}}$ in the unbiased case.}}\label{fig:POVMBias} 
\end{figure}

This formulation captures the unsharpness and bias of the probe measurement \(\mathcal{M}_a\) through a combined transformation that performs both skewing (shearing) and scaling (squeezing) on the original circle. Here, \(\delta\) is the shear factor, quantifying the measurement bias.
Next we reverse the order of POVM and projective measurement; now the probe (first) measurement is the POVM $\m M_{a}(\gamma)=E_{+\vert a}(\gamma)-E_{-\vert a}(\gamma)$ and the target (second) measurement is projective, given by $\m M_{b}(\theta)=\cos\theta\sigma_x+\sin\theta\sigma_z$. As seen from Fig.\,\ref{fig:POVMBias} the (unbiased) unsharp measurement disturbs the second measurement less compared to a projective measurement, which results in an elliptic shape, as defined in Eq. (\ref{ellipse}) with $s<\vert \vec a\vert$, of the correlation-disturbance tradeoff, plotted in red and purple for interaction strengths $\gamma=0.5$ and  $\gamma=0.75$, respectively. Finally, the effect of a biased POVM is studied, where according to Eq. (\ref{ellipse}) in addition the elliptic shape  of the correlation-disturbance relation a tilt and shearing, represented by shear factor $\delta$ occurs, which is plotted in Fig.\,\ref{fig:POVMBias} for $\gamma=0,5,\,a_0=0.5$ and $\gamma=0,75,\,a_0=0.25$ as red and purple squares, respectively. 

%%%%%%%%%%%%%%%%%%%
%{\it \textcolor{blue}{High-dimensional case.---}}
{\it Optical quantum detector noise model}.---
%%%%%%%%%%%%%%%%%%
Let us examine the covariant $\m{CD}$-tradeoff for optical quantum detectors. Single-photon detectors are an important type of quantum detector \cite{Hadfield2009}. These binary (on-off) detectors register a detection outcome (click or on) upon the arrival of any number of photons, whereas no detection (no click or off) occurs otherwise. Dark count (probability of the click in the absence of a photon, characterized by parameter $\nu$) and detector efficiency (probability of a click upon photon arrival, characterized by parameter $\eta$) are critical characteristics of single-photon detectors. The pertinent POVM elements, accounting for these primary noise factors are expressed as $E_{\rm off}=e^{-\nu}\sum_{n=0}^\infty (1-\eta)^n\proj{n}$ and $E_{\rm on}=\mathbbm 1-E_{\rm off}$. The aforementioned noise characteristics significantly impact the security and efficiency of numerous quantum communication protocols due to the challenge of differentiating between actual and spurious events, complicating fair sampling. Thus, accurately and reliably assessing intrinsic detector readout noise independent of unitary or transmission channel noises is essential. \Stephan{See Supplemental Material \cite{Supp} Sec. II for details and schematic illustration of the optical arrangement.} Single photons are prepared using well-established techniques to examine joint measurements, where the initial measurement uses a polarizing beam splitter to evaluate photon polarization encoded in two distinct paths, and the second measurement determines its presence in a specific mode via the on-off detector. We can similarly define correlations as joint probabilities $p(\alpha,\beta)$ with $\alpha\in \{\rm H\equiv-1, V\equiv+1\}$ and $\beta\in\{\rm off\equiv +1, on\equiv-1\}$ and disturbance. We can directly estimate the noise parameters $\eta$ and $\nu$ by examining two cases: (1) when the reference measurement determining the polarization mode is sharp, and (2) when the measurement is fully biased, allowing all photons to pass through. In case (1), according to the $\m {CD}$-tradeoff the maximum disturbance is given by $\m D^{(1)}=e^{-\nu}\eta$, while $\m C^{(1)}=0$. Conversely, in situation (2), we have $\m C^{(2)}=e^{-\nu}(2-\eta)-1$ and $\m D^{(2)}=0$. Consequently, the characteristic noise parameters $\eta$ and $\nu$ can be inferred solely from $\m D^{(1)}$ and $C^{(2)}$, eliminating the need for detector tomography \cite{Lundeen2009,Eisaman2011,Zhang2020,Orsucci2020,Hlousek2021}. Furthermore, a channel may be introduced between measurements to compare outcomes and assess channel loss. It is worth pointing out that this strategy is well-tailored for photon detectors where the source of the noise is from the intrinsic part of the detector rather than the random unitaries,e.g., polarization rotation.

%%%%%%%%%%%%%%%%%%%
{\it Conclusion and outlook.---}
%%%%%%%%%%%%%%%%%%
The correlation 
$\m C$ and the disturbance $\m D$
induced by quantum measurements are not independent quantities; they are mutually related in a nontrivial way. This relationship is expressed as covariant symmetry interplay between the two quantities. Exploiting these symmetries provides a robust tool for estimating characteristic parameters of measurement devices, such as sharpness and bias with  practical application in high-dimension settings. Our experimental implementation efficiently demonstrates our theoretical findings, covariant $\m {CD}$-tradeoff.

At a fundamental level, we propose that optimal configurations of disturbance, as a way of transmitting information, and correlation, as a measure of predictability, could outperform or even rule out certain classical (hidden variable) models. Associating a classical (noncontextuality) inequality within our scheme offers a new perspective on the extent of nonclassicality implied by quantum uncertainties \cite{Catani} which also aligns with schemes of two-time Leggett-Garg macroscopic realism \cite{TwoLG}, sequential random access codes and self-testing of measurement instruments \cite{Mohan2019, Miklin, Wagner, Anwer, Chen}. Consequently, our scheme could be effectively adapted to optical-based communication tasks where quantum advantage over classical counterparts can be explored.

\begin{acknowledgments}

\textcolor{black}{We thank Marcus Huber and Giuseppe Vitagliano for fruitful discussions and useful
comments. }This work was supported by the Austrian Science Fund (FWF) Project No. P 34239.

\end{acknowledgments}

\appendix
%\section{Supplementary Material}

\vspace{8mm}

\quad \quad \quad\quad \quad\quad \quad\quad {\textbf{APPENDICES}}

\vspace{-8mm}

\section*{}
This Supplemental Material presents the theoretical derivation of the correlation-disturbance tradeoff relation, a practical application of our scheme in terms of a single-photon detector, and details of the experimental procedures, including additional results.


\section{Theoretical Details}

\subsection{An explicit derivations \& Proof of the inequality}

First, let us define the disturbance operator \(\hat{d}\) and the disturbance \(\mathcal{D}\) as:
\begin{align}
\hat{d} = \mathcal{M}_b - \mathcal{I}^*_a(\mathcal{M}_b), \quad \mathcal{D} = \langle \hat{d} \rangle_\rho,
\end{align}
where \(\mathcal{I}^*_a(\mathcal{M}_b)\) is the action of the Lüders channel on \(\mathcal{M}_b\), given by:
\begin{align}
\mathcal{I}^*_a(\mathcal{M}_b) = \sum_\alpha \mathcal{I}^*_\alpha(\mathcal{M}_b) = \sum_\alpha E_\alpha^{\frac{1}{2}} \mathcal{M}_b E_\alpha^{\frac{1}{2}}.
\end{align}

A useful trick is to decompose the Lüders channel into two terms:
\begin{align}
\mathcal{I}^*_\alpha(\mathcal{M}_b) = \frac{1}{2}\{E_\alpha, \mathcal{M}_b\} - \mathcal{L}^*_{\alpha}(\mathcal{M}_b),
\end{align}
where the second term, \(\mathcal{L}^*_{\alpha}(\mathcal{M}_b)\), often referred to as the dissipator, accounts for the disturbance:
\begin{align}
\mathcal{L}^*_{\alpha}(\mathcal{M}_b) &= \frac{1}{2} \Big[E_\alpha^{\frac{1}{2}}, [E_\alpha^{\frac{1}{2}}, \mathcal{M}_b]\Big]= \Big[E_\alpha^{\frac{1}{2}}, [E_\alpha^{\frac{1}{2}}, E_{+|b}]\Big].
\end{align}
Thus, we can express the disturbance operator \(\hat{d}\) as:
\begin{align}
\hat{d} = \mathcal{L}^*_{+|a}(\mathcal{M}_b) + \mathcal{L}^*_{-|a}(\mathcal{M}_b)=2\Big[E_\alpha^{\frac{1}{2}}, [E_\alpha^{\frac{1}{2}}, E_{+|b}]\Big],
\end{align}

The correlation, denoted by \(\mathcal{C} = \mathrm{tr}(\rho \hat{C})\), can be written as:
\begin{align}
\hat{C} = \sum_\alpha \alpha \mathcal{I}^*_\alpha(\mathcal{M}_b),
\end{align}
which can be expanded as:
\begin{align}
\label{CorrOp}
\hat{C} = \frac{1}{2} \{\mathcal{M}_a, \mathcal{M}_b\} - \Big(\mathcal{L}^*_{+|a}(\mathcal{M}_b) - \mathcal{L}^*_{-|a}(\mathcal{M}_b)\Big).
\end{align}
Note that the above derivations are general and apply to arbitrary dichotomic measurements in $d$-dimensional Hilbert space. This includes some interesting examples, such as parity measurements, on-off photon detectors, and collective $n$-qubit measurements, i.e. $\m M=U\sigma_z^{\otimes n} U^\dag=\Pi_+ -\Pi_-$.

\subsection{General qubit measurement}

The spectral decomposition of the Positive Operator-Valued Measure (POVM) elements for a general dichotomic measurement is given by
\begin{align}
E_+ = \sum_{j=1}^2 p_j \proj{j}, \quad E_- = \mathbbm{1} - E_+ = \sum_{j=1}^2 q_j \proj{j},  
\end{align}
where \( q_j = 1 - p_j \). This describes how \(E_+\) and \(E_-\) are built from projectors \( \proj{j} \) with probabilities \(p_j\) and \(q_j\), ensuring \( E_+ + E_- = \mathbbm{1} \). This leads to the relation:
\begin{align}
E_{\pm|b} = \frac{\mathbbm{1} \pm \mathcal{M}_b}{2}, \quad \mathcal{M}_b = b_0 \mathbbm{1} + |\vec{b}| \hat{b} \cdot \vec{\sigma},
\end{align}
where \( \mathcal{M}_b \) is parameterized by the four-vector \( (b_0, \vec{b}) \), and similarly for \( \mathcal{M}_a \) with \( (a_0, \vec{a}) \).

A useful form for general two-outcome POVMs, such as \( \mathcal{M}_a \) and \( \mathcal{M}_b \), is:
\begin{align}
E_{\pm|b} = |\vec{b}| \Pi_{\pm|b} + (1 - |\vec{b}|) N_{\pm|b}, 
\end{align}
where \( E_{\pm|b} \) is decomposed into a projection term \( \Pi_{\pm|b} \) and a noise term \( N_{\pm|b} \), known as randomizing the measurement.

Finally, the joint probability for two outcomes \( (+, +) \) is:
\begin{align}
p(+,+) = \mathrm{tr} \left[ \mathcal{I}_{\alpha|a}(\rho) E_{\beta|b} \right],
\end{align}
where \( \mathcal{I}_{\alpha|a}(\rho) \) is the post-measurement state conditioned on outcome \( \alpha \) for measurement \( a \), and \( E_{\beta|b} \) is the POVM element for outcome \( \beta \) of measurement \( b \).

Now, more explicitly, for a general two-outcome qubit POVM:
\begin{align}
\hat{d} = |\vec{b}|\Big(\mathcal{L}^*_{+|a}(\hat{b} \cdot \vec{\sigma}) + \mathcal{L}^*_{-|a}(\hat{b} \cdot \vec{\sigma})\Big).
\end{align}
The following relation simplifies the derivation:
\begin{align}
(\vec{a} \cdot \vec{\sigma})(\vec{b} \cdot \vec{\sigma}) = \vec{a} \cdot \vec{b} \, \mathbbm{1} + i (\vec{a} \times \vec{b}) \cdot \sigma.
\end{align}
Using this, the dissipator \(\mathcal{L}^*_\alpha(\mathcal{M}_b)\) can be simplified as:
\begin{align}
\mathcal{L}^*_\alpha(\mathcal{M}_b) = \frac{a_\alpha s_\alpha |\vec{b}|}{2}\Big(\hat{a} \times (\hat{a} \times \hat{b})\Big) \cdot \sigma,
\end{align}
where \(a_\pm = 1 \pm a_0\) and \(s_\pm = 1 - \sqrt{1 - \frac{|\vec{a}|^2}{a_\pm^2}}\). The term \(s\) is then:
\begin{align}
s = \frac{1}{2}(a_+ s_+ + a_- s_-) = 1 - \frac{1}{2}(u_+ + u_-),
\end{align}
where, as defined earlier, \(u_\alpha = \sqrt{a_\alpha^2 - |\vec{a}|^2}\). For unbiased POVMs, we have \(a_{\pm} = 1 - \sqrt{1 - |\vec{a}|^2}\). Therefore, the disturbance \(\mathcal{D}\) for the optimal state is:
\begin{align}
\mathcal{D} = s |\vec{b}| \sin \theta.
\end{align}

We also need to simplify the following expression:
\begin{align}
\frac{1}{2}(a_+ s_+ - a_- s_-) = a_0 + \delta,
\end{align}
where \(\delta = \frac{1}{2}(u_+ - u_-)\) represents the measurement bias.

The correlation \eqref{CorrOp} for qubit POVM read:
\begin{align}
\hat{C} = \frac{1}{2} \{\mathcal{M}_a, \mathcal{M}_b\} - |\vec{b}|\Big(\mathcal{L}^*_{+|a}(\hat{b} \cdot \vec{\sigma}) - \mathcal{L}^*_{-|a}(\hat{b} \cdot \vec{\sigma})\Big).
\end{align}

Using these relations, the correlation for the optimal prepared state becomes:
\begin{align}
\mathcal{C} = a_0 b_0 + |\vec{a}||\vec{b}| \cos \theta + \delta |\vec{b}| \sin \theta.
\end{align}

The trade-off between correlation \(\mathcal{C}\) and disturbance \(\mathcal{D}\) is given by the following parametric equations:
\begin{align}\label{eq:ell}
\left[\begin{array}{c} \mathcal{C} \\ \mathcal{D} \end{array}\right] = \left[\begin{array}{c} a_0 b_0 \\ 0 \end{array}\right] + \left[\begin{array}{cc} |\vec{a}| & \delta \\ 0 & s \end{array}\right] \left[\begin{array}{c} |\vec{b}| \cos \theta \\ |\vec{b}| \sin \theta \end{array}\right].
\end{align}

This formulation captures the unsharpness and bias of the probe measurement \(\mathcal{M}_a\) through a combined transformation that performs both skewing (shearing) and scaling (squeezing) on the original circle. Here, \(\delta\) is the shear factor, quantifying the measurement bias.

We can proceed with further simplifications. The expression:
\begin{align}
|\vec{a}||\vec{b}|\cos \theta + \delta |\vec{b}|\sin \theta
\end{align}
can be simplified into a single cosine term:
\begin{align}
R \cos(\theta - \phi),
\end{align}
where
\begin{align}
R = |\vec{b}|\sqrt{|\vec{a}|^2 + \delta^2} \quad \text{(the amplitude)},
\end{align}
\begin{align}
\tan \phi = \frac{\delta}{|\vec{a}|} \quad \text{(the phase shift)}.
\end{align}
This derivation resembles elliptical polarization.

Correlation and disturbance can generally be expressed as
\begin{equation}
    \mathcal{C} = \lambda u_0 + (1 - \lambda)v, \quad \mathcal{D} = \lambda u_1,
\end{equation}
where the square of \(\mathcal{C}\) and \(\mathcal{D}\) are given by:
\begin{align}
    \mathcal{C}^2 &= \lambda^2 u_0^2 + 2 \lambda (1 - \lambda) u_0 v + (1 - \lambda)^2 v^2, \\
    \mathcal{D}^2 &= \lambda^2 u_1^2.
\end{align}
Then,
\begin{align}
    \mathcal{C}^2 + \mathcal{D}^2 &\leq \lambda^2 \| u \|^2 + 2 \lambda (1 - \lambda) u_0 v + (1 - \lambda)^2 v^2 \\
    &\leq \Big(\lambda \| u \| + (1 - \lambda) v \Big)^2 \leq 1,
\end{align}
where we use \(\| u \|^2 = u_0^2 + u_1^2\).

The inequality holds because \(u_0 \leq \sqrt{1 - u_1^2}\), implying \(u_0 \leq \| u \| \leq 1\).

\subsection{Generalization to two-outcome measurements in $d$-dimensional Hilbert space}

The spectral decomposition of the POVM elements for a general dichotomic measurement in a \(d\)-dimensional Hilbert space is given by
\begin{align}
E_+ = \sum_{j=1}^d p_j \proj{j}, \quad E_- = \mathbbm{1} - E_+ = \sum_{j=1}^d q_j \proj{j},  
\end{align}
where \( q_j = 1 - p_j \). 

A specific example is a randomized measurement that decomposes as a projection plus a noise term:
\begin{align}
E_+ = \gamma \Pi_+ + (1 - \gamma) N_+, \quad E_- = \mathbbm{1} - E_+,
\end{align}
where \(\gamma\) quantifies the measurement’s sharpness or precision. The noise term does not have to be proportional to the identity. The only fact about it is that it does not contribute to disturbance.

In the case where the first measurement is sharp and the second measurement follows the above form, the disturbance operator \(\hat{d}\) is given by
\begin{align}
    \hat{d} &= \gamma \sum_\alpha \Big(\{\Pi_\alpha, \Pi_{+|b}\} - 2 \Pi_\alpha \Pi_{+|b} \Pi_\alpha \Big) \\
    &= 2 \gamma \Big(\{\Pi_{+|a}, \Pi_{+|b}\} - 2 \Pi_{+|a} \Pi_{+|b} \Pi_{+|a}\Big).
\end{align}

Operator $\{\Pi_{+|a}, \Pi_{+|b}\} - 2 \Pi_{+|a} \Pi_{+|b} \Pi_{+|a}$ is a traceless rank-two matrix, meaning that its eigenvalues are \(\pm \lambda\) and \(0\).
Thus, we can express the spectral decomposition of \(\hat{d}\) as
\begin{equation}
    \hat{d} = \gamma\Big[\lambda \proj{\psi_+} - \lambda \proj{\psi_-}\Big],
\end{equation}
where
\begin{equation}
    \lambda = 2 \sqrt{(1 - c^2) c^2} \ \ \ , \quad c^2 = \tr\big(\Pi_{+|a} \Pi_{+|b}\big).
\end{equation}
The maximum value of disturbance,  \(\lambda=1\), is reached when \( c^2 = \frac{1}{2}\). Note that, 
\begin{equation}
    \Pi_{+|a}=\proj{v_a} \ \ \ \ , \ \ \ \  \Pi_{+|b}=\proj{v_b}
\end{equation}
and $|\bra{\psi_+}v_a\rangle |=\frac{1}{\sqrt 2}$ and consequently, we obtain
\begin{equation}
\mean{\m M_a}=0.
\end{equation}
Note that the optimal state is such that the value of the disturbance solely depends on the overlap between the projectors of the corresponding measurements.
\begin{equation}
    \mathcal{D} = \bra{\psi_+} \hat{d} \ket{\psi_+} = 2 \gamma\sqrt{(1 - c^2) c^2}
\end{equation}
and for the correlation
\begin{equation}
    \mathcal{C} = \bra{\psi_+} \hat{C} \ket{\psi_+} =\gamma (2c^2 - 1).
\end{equation}
This yields the relation
\begin{equation}
    \mathcal{C}^2 + \mathcal{D}^2 = \gamma^2,
\end{equation}
satisfying a circle in the \(\mathcal{C}\)-\(\mathcal{D}\) plane.
Further, we can directly estimate the Bloch length of the second measurement as \( |\vec{b}| = \gamma \sqrt{d - 1} \). This length is chosen such that for a basis operator, the orthogonality condition \(\tr(\sigma_i \sigma_j) = d \delta_{ij}\) holds and thus $0\leq|\vec b|\leq\sqrt{d-1}$

%%%%%%%%%%%%
\section{Estimating Readout Noise of a Single-Photon Detector}

In this section, we aim to show that the covariance between correlation and disturbance can be used to estimate the characteristic noise of a single-photon detector — often called an on-off detector — which registers a click when the input state is non-vacuum. Such detectors suffer from two main sources of noise: photon loss and dark counts, characterized by the parameters $\eta$ and $\nu$, respectively.

\begin{figure}[h!]
\includegraphics[width=9cm]{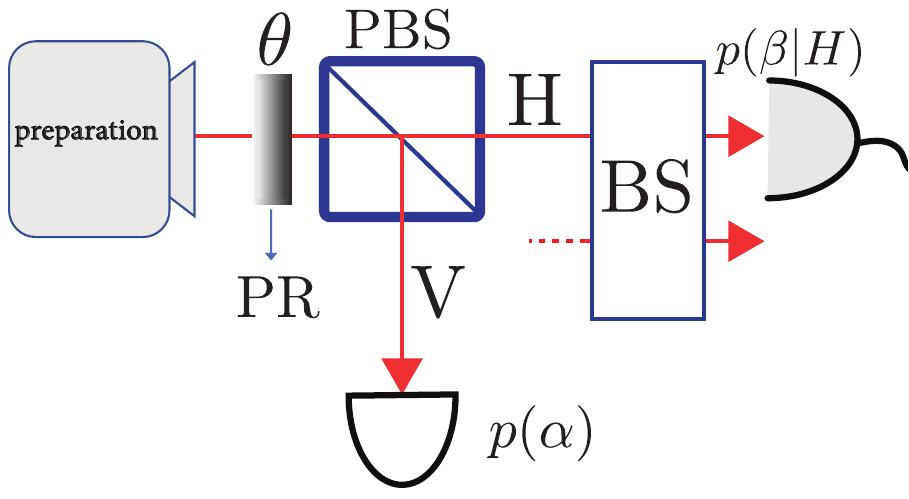}	
\caption{The source emits single photons with a known total number. A variable polarization rotator (PR) transforms the polarization of the photons. The horizontal and vertical components are then encoded in two separate path modes via a polarizing beam splitter (PBS), in front of which an ideal detector is placed. This constitutes the (first) reference measurement, denoted by $\m M_a$. Photons projected into the orthogonal path are directed to the second (probe) measurement, $\m M_b$, whose setting can be varied using a standard beam splitter.}
\label{Detector} 
\end{figure}

The experimental setup that realizes the correlation and disturbance is illustrated in Fig.~\ref{Detector}. The sequential measurement consists of two distinct dichotomic measurements. The first measures the polarization degree of freedom, which is encoded in the spatial modes via a polarizing beam splitter (PBS).

The polarization measurement and photon detection can be viewed as a joint measurement. Let us now show how to evaluate the probabilities explicitly.
Our joint measurement consists of a perfect detector (shown in white) used to measure the polarization of the input photon. Upon a no-click (negative) result, we infer that the photon exists in the other polarization mode, which is directed to the on-off detector. The detector POVM elements, accounting for these primary noise factors are expressed as 
\begin{align}
&E_{\rm off}=e^{-\nu}\sum_{n=0}^\infty (1-\eta)^n\proj{n},\nonumber\\
&E_{\rm on}=\mathbbm 1-E_{\rm off}.
\end{align}

The joint probabilities are given by:
\begin{align}\label{eq:POVMPhot}
    p({\rm H, on}) &= p({\rm H})\, \tr(\rho_H E_{\rm on}), \nonumber\\
    p({\rm H, off}) &= p({\rm H})\, \tr(\rho_H E_{\rm off}),\nonumber \\
    p({\rm V, on}) &= p({\rm V})\, \tr(\rho_V E_{\rm on}),\nonumber \\
    p({\rm V, off}) &= p({\rm V})\, \tr(\rho_V E_{\rm off}),
\end{align}
from which we compute the correlation:
\begin{equation}
    \m C = 2\big( p({\rm H, on}) + p({\rm V, off}) \big) - 1.
\end{equation}

The disturbance can be obtained by comparing $p(\rm off)$ and $\tilde{p}(\rm off)$, where
\begin{equation}
    \tilde{p}({\rm off}) = p({\rm H, off}) + p({\rm V, off}),
\end{equation}
which corresponds to the case where we detect photons regardless of their polarization. In this scenario, the updated state sent to the second measurement is 
\[
\m I_{a}(\rho) = \frac{1}{2} \proj{1_H} + \frac{1}{2} \proj{0}.
\]
Recall that the disturbance is given by
\[
\m D = 2 \left| p({\rm off}) - \tilde{p}({\rm off}) \right|.
\]
Thus, when estimating disturbance, it is not necessary to read out the photon polarization.

A clarifying remark is in order: the polarization-mode measurement and the corresponding post-measurement state are realized via the “no-click” event. The no-click is interpreted as a measurement outcome ($H$), indicating the photon is in the horizontal polarization, just as a click corresponds to the $V$ outcome. We can, therefore, fairly evaluate $p(\rm H,on)$ and $p(\rm H,off)$. Likewise the other joint probabilities are evaluated by making negative measurement (no-click) for outcome ($V$).

Let us now examine the covariance-based trade-off relation in different scenarios. First, consider a simple configuration where the polarization rotator (PR) and the beam splitter (BS) are removed. The optimal input state that maximizes disturbance is:
\begin{equation}
    \ket{\psi_{\rm opt}} = \ket{1_{\rm D}} = \frac{1}{\sqrt{2}}\left(\ket{1}_H \ket{0}_V + \ket{0}_H \ket{1}_V\right),
\end{equation}
yielding:
\begin{equation}
\label{D1}
    \m D^{(1)} = e^{-\nu} \eta, \quad \m C^{(1)} = 0.
\end{equation}

this capture the sharpness of the quantum detector.
In another configuration, we introduce a fully biased reference measurement, which passes all photons as if they are horizontally polarized. This results in:
\begin{equation}
\label{C2}
    \m D^{(2)} = 0, \quad \m C^{(2)} = e^{-\nu}(2 - \eta) - 1.
\end{equation}
Note that $\m C^{(2)}$ reflects only the bias of the second measurement in this case. For instance, it vanishes when the measurement is sharp (i.e., $\eta = 1$ and $\nu = 0$), meaning it is unbiased. Using Eqs.~\eqref{D1} and \eqref{C2}, we can estimate the noise parameters $\eta$ and $\nu$ as:
\begin{equation}
    \eta = \frac{2\m D^{(1)}}{\m C^{(2)} + \m D^{(1)} + 1}, \quad 
    \nu = -\ln \left(\frac{\m C^{(2)} + \m D^{(1)} + 1}{\m D^{(1)}}\right).
\end{equation}
These are the two parameters that reconstruct the POVM associated with the detector from Eq. (\ref{eq:POVMPhot}).

%%%%%%%%%%%%%%%%%%%%%
\section{Experimental Details and Additional Results}
%%%%%%%%%%%%%%%%%%%%%
\subsection{Neutron Optical Setup}
The experiment was carried out at the polarimeter instrument {\it{NepTUn (NEutron Polarimeter TU wieN)}}, located at the tangential beam port of the 250\,kW TRIGA Mark II  research reactor at the Atominstitut - TU Wien, in Vienna, Austria. As in a number of previous experiments the qubit system studied in the present experiment is represented by the spin, a {\it{two-state system}}, of the spin-1/2 particle neutron, where $\mathcal S_i=\frac{\hbar}{2}\,\sigma_i$, with $i=x,y,z$. An unpolarized thermal neutron beam is monochromatized via Bragg-reflection from a pyrolytic graphite crystal, having a mean wavelength of 2.02 \AA\, and spectral width $\Delta\lambda/\lambda=0.015$. Next the beam is spin polarized up to $\sim99\,\%$ in $+z$-direction by reflection from a bent Co-Ti supermirror array thus yielding spin state $\vert+z\rangle$. In order to manipulate the neutron's spin state magnetic fields are applied which interact with the neutron via the Zeeman Hamiltonian $\mu\vec\sigma\vec B$. The action of a static magnetic field pointing in direction $\vec n_B$ on the neutron's spin state is described by the unitary transformation $U_R=e^{\textrm{i}\alpha \vec n_B\vec\sigma}$, with $\alpha=\mu\vert B\vert T/\hbar$ and $\vec\sigma=(\sigma_x,\sigma_y,\sigma_z)^T$, where $\sigma_{x,y,z}$ are the Pauli spin operators. Here, $\mu$ denotes the magnetic moment of the neutron, $\vert B\vert$ the modulus of the magnetic field strength, and $T$ the time of flight through the field region. Thus, the static magnetic field induces a rotation of the neutron's polarization vector, i.e., the neutron's Bloch vector, around the axis $\vec n_B$ of the magnetic field, with frequency $\omega_{\rm L}= \vert 2 \mu B/\hbar\vert$, which is referred to as Larmor precession. For thermal neutrons, effects of the magnetic field on the spatial part of the neutron's wave function (longitudinal Stern - Gerlach effect) are negligible. To prevent depolarization by stray fields, a guide field $B^{\rm{GF}}_z$ pointing in the positive $z$-direction having a magnitude of 13 Gauss, from coils in Helmholtz configuration, is applied along the entire setup.

For the preparation of arbitrary input states $\vert\psi_{\rm in}\rangle$ a so-called spin rotator coil DC-1 is used which produces a static magnetic field $B_y$ pointing in $+y$-direction. The Bloch vector of the incident neutrons thus experiences Larmor precession around the $y$-axis while passing through the coil. By setting the coil's field strength the Bloch vector's polar angle is set to the required value.

For the projective measurement of a spin observable $\m M_a=2 M_{+|a}-\mathbbm 1$ its projection operators $M_{+,1|a}$ have to be realized in the actual experiment. We illustrate this procedure for the case $\m M_a=\sigma_x$ (and as well for $\m M_a=\cos\theta\sigma_x+\sin\theta\sigma_z$) in Fig.\,2 of the main text. Therefore, a second spin turner coil, denoted as  DC-2, with a magnetic field pointing in $+y$-direction is utilized. The field strength of DC-2 is chosen such that exactly the $\vert+x\rangle$ ($\vert+\theta\rangle$) component of the initially prepared state is rotated onto $\vert+z\rangle$. Next, another supermirror array (first analyzer) only allows neutrons in $\vert+z\rangle$ spin state to pass thus acting as projector $\vert +z\rangle\langle +z\vert$. From the $\vert+z\rangle$ state exiting the supermirror $\vert+z\rangle$ is generated by using another spin turner coil (DC-3). Analogously to the initial state preparation with DC-1, DC-3's field strength adjusts the polar angle of the neutron's Bloch vector to complete the projective measurement, with {\it{post-measurement state}} being an  eigenstate of $\m M_a$ (unsharp measurements are discussed in the next Section). 

The subsequent measurement of $\m M_b=2 M_{+|b}-\mathbbm 1$ is performed in similar manner with spin turner coil DC-4 and a second supermirror analyzer. A final preparation of the eigenstates of $\m M_b=2 M_{+|b}-\mathbbm 1$ is not required since the BF$_3$ counting detector is insensitive to spin states. To compensate random fluctuations of the reactor's neutron flux an additional $^3$He monitor detector, mounted in front of the actual setup, is used for normalization of the count rate.

\subsection{Correlation and Distrubance}

\begin{figure}[b!]
\includegraphics[width=8.5 cm]{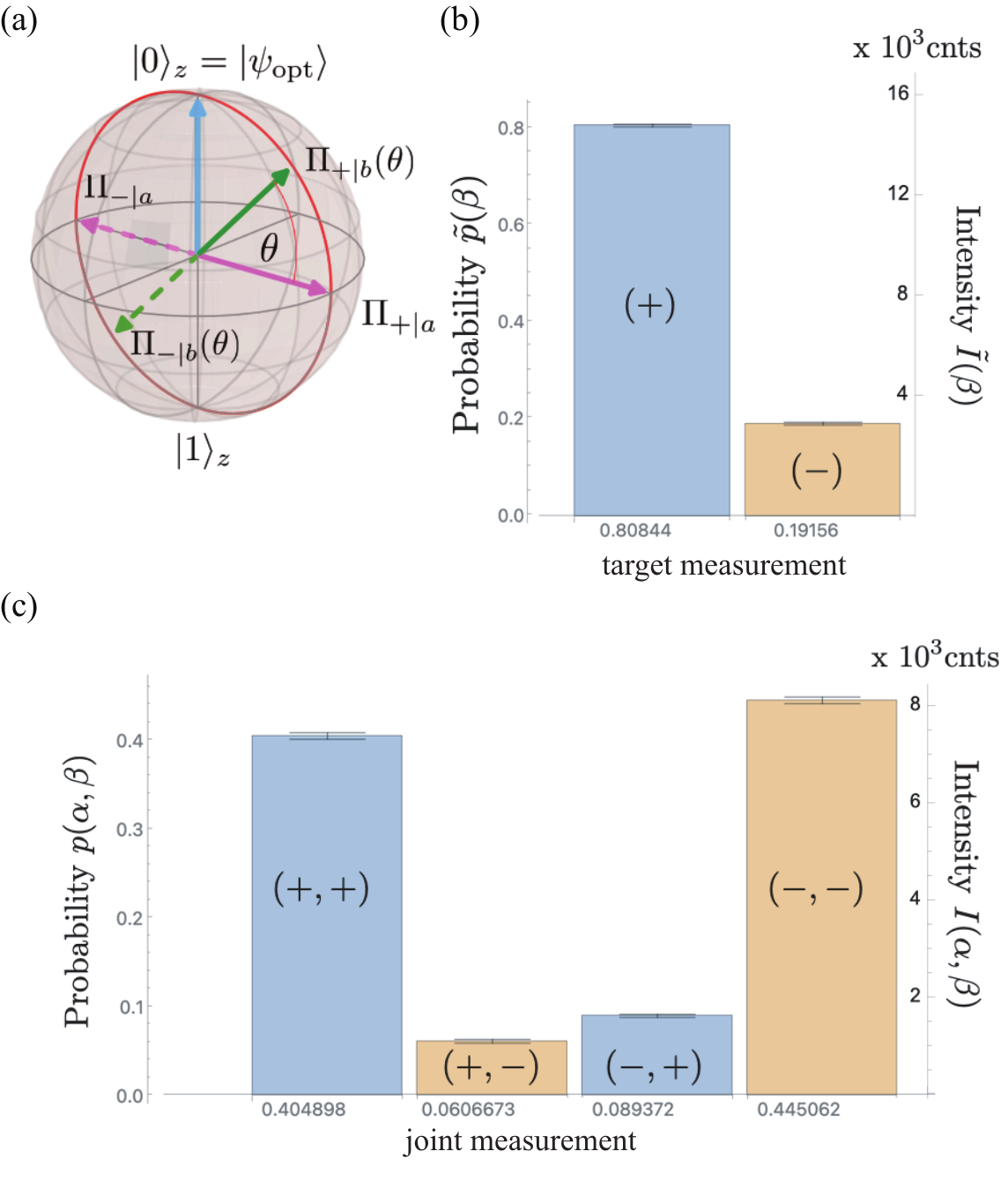}	
\caption{Measured intensities for measurement setting with (fixed) $\m M_{a}=\sigma_x$ and varying $\m M_{b}(\theta)=1/\sqrt 2\, \sigma_x+1/\sqrt 2\,\sigma_z$ ($\theta=\pi/4$). (a) Bloch sphere representation (b) target measurement alone (with $\m M_{a}$ off) and (c) joint measurement. Error bars represent $\pm$ one standard deviation.
}\label{fig:Intensities} 
\end{figure}

The successive measurement of $\m M_a$ and $\m M_b$ (sharp or unsharp) finally result in four recorded intensities denoted as $I_{++}, I_{+-}, I_{-+}$, and $I_{--}$, where the first index denotes the measurement orientation of the $\m M_a$ measurement and second for $\m M_b$. Using these intensities the correlation $\m C=2p(\alpha=\beta)-1$ is explicitly given by
\begin{equation}
\mathcal C=2\,\frac{I(+,+)+I(-,-)}{I(+,+)+I(+,-)+I(-,+)+I(-,-)}-1.
\end{equation}
Furthermore, we can evaluate the disturbance given by $\m D=2|p(+|b)-\tilde p(+|b)|$ as
\begin{align}
    \mathcal D=&2\Bigg\vert\frac{\tilde I(+)}{\tilde I(+)+\tilde I(-)}\nonumber\\
    &-\frac{I(+,+)+I(-,+)}{I(+,+)+I(+,-)+I(-,+)+I(-,-)} \Bigg\vert,
\end{align}
where $I^\prime_\pm$ accounts for the intensities obtained in a {\it{target}} measurement alone, that is a measurement with apparatus $\m M_a$ switched off, and therefore only two results, denoted as $I_\pm^\prime$, are obtained. See Fig.\,\ref{fig:Intensities} for the measured intensities $I(\alpha,\beta)$ from the joint and $\tilde I(\beta)$ from the target measurement.

%%%%%%%%%%%%%%%%%

%%%%%%%%%%%%%%%%
%%%%%%%%%%%%%%%%%

\subsection{Projective Measurements}

%%%%%%%%%%%%%%%%
\begin{figure}[b!]
\includegraphics[width=8.5 cm]{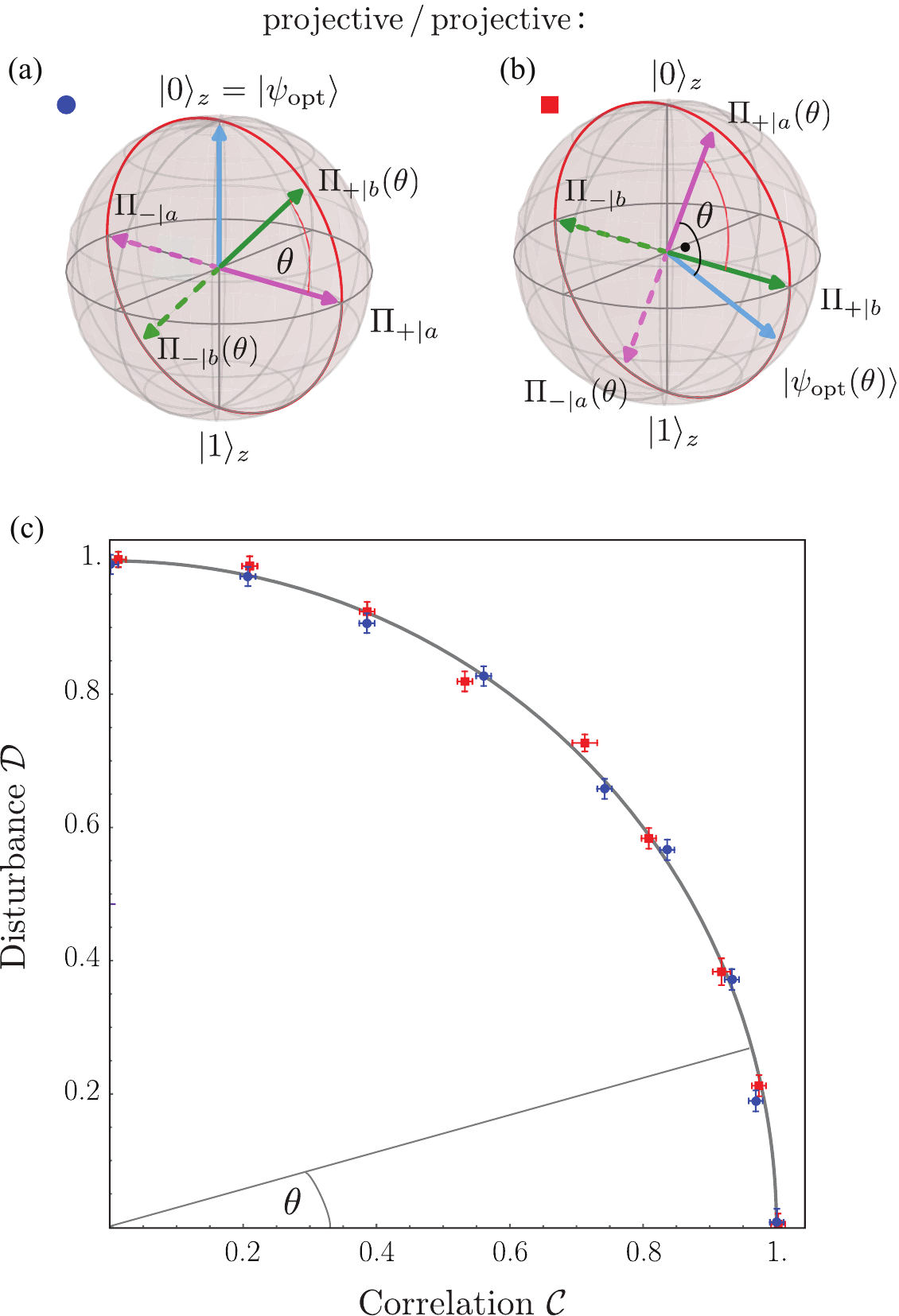}	
\caption{Successive projective measurements. (a) Bloch sphere representation for fixed $\m M_{a}=\sigma_x$ and varying $\m M_{b}(\theta)=\cos\theta\sigma_x+\sin\theta\sigma_z$ and (b)  fixed $\m M_{b}=\sigma_x$ and varying $\m M_{a}(\theta)=\cos\theta\sigma_x+\sin\theta\sigma_z$ and corresponding optimal state $\ket{\psi_{\rm{opt}}(\theta)}$. (c) Plot of tradeoff relation between  disturbance $\m D$ and correlation $\m C$.
}\label{fig:CorrVsDist} 
\end{figure}
%%%%%%%%%%%%%%%%%

First, we implement in our setup the special example in which both measurements are projective (labeled as $\Pi_{\pm\vert i}$). The first measurement is fixed to be $\m M_a=\Pi_{+\vert a}-\Pi_{-\vert a}=\sigma_x$, whereas the second subsequent measurement is set to $\m M_b=\Pi_{+\vert b}(\theta)-\Pi_{-\vert b}(\theta)=\cos\theta\sigma_x+\sin\theta\sigma_z$, with experimentally adjustable parameter $\theta$, controlled by the current in DC-4, generating the appropriate magnetic field pointing in $+y$-direction. The disturbance and correlation are given in theory by $\m D=\sin\theta\mean{\sigma_z}_\rho$ and $\m C=\cos\theta$, respectively. The optimal input state is $\ket{\psi_{\rm opt}}=\ket{0}_z$ for all settings of the controlled parameter $\theta$.  Figure\,\ref{fig:CorrVsDist}\,(a) shows a plot of the disturbance $\m D$ versus the correlation $\m C$; for $\theta=0$, accounting for a compatible measurement setting of $\m M_a=\sigma_x$ and $\m M_b=\sigma_x$, we consequently observe zero disturbance $\m D$ and maximal correlation $\m C=1$. For the other extreme case $\theta=\pi/2$ we have maximal disturbance $\m D=1$ and no more correlations. For all other values $0<\theta<\pi/2$ we observe a tradeoff between $\m D$ and correlation  $\m C$, always fulfilling the tight relation $\m C^2+\m D^2=1$, demonstrating our main results as predicted in Eqs. (5) and (6) from the main text.

Next we investigate the case $\m M_a=\Pi_{+\vert a}(\theta)-\Pi_{-\vert a}(\theta)=\vec b\cdot\vec\sigma =\cos\theta\sigma_x+\sin\theta\sigma_z$ and $\m M_b=\Pi_{+\vert b}-\Pi_{-\vert b}=\vec b\cdot\vec\sigma=\sigma_x$ with associate Bloch vector $\vec a=(\cos\theta,0,\sin\theta)$ and $\vec b=(1,0,0)$, respectively. Now the optimal initial state $\vert\psi_{\rm{opt}}\rangle$ is no longer independent of the control parameter $\theta$ (see Supplementary Material Sec II\,D for the experimental search of optimal initial state), \aa{ ensuring their operation within the expected parameters entitled by the covariant symmetry between correlation and disturbance.}
%To verify Eq.(\ref{eq:optstate}) an experimental search of the optimal initial $\vert\psi_{\rm{opt}}\rangle$ is performed, where the polar angle $\phi$ of an arbitrary initial state $\vert\psi_{\rm{in}}(\phi)\rangle$ is varied systematically (see Supplementary Material experimental details). 
The plot of disturbance $\m D$ versus the correlation $\m C$ in Fig.\,\ref{fig:CorrVsDist}\, (b) shows a similar behavior as before where the $\m M_a$ measurement was fixed; for $\theta=0$ (compatible measurement setting of $\m M_a=\sigma_x$ and $\m M_b=\sigma_x$) zero disturbance $\m D$  and maximal correlation $\m C=1$ are observed. For the other extreme case $\theta=\pi/2$ we have maximal disturbance $\m D=1$ and no more correlations. \aa{For all other values $0<\theta<\pi/2$ the covariance between $\m D=\cos\theta$ and correlation  $\m C=\sin\theta$ fulfills} the tight relation $\m C^2+\m D^2=1$, as a special case of Eq. (6) demonstrating our main result.

\subsubsection{Search for optimal state}

In case $\m M_a=\vec a\cdot\vec\sigma =\cos\theta\sigma_x+\sin\theta\sigma_z$ and $\m M_b=\vec b\cdot\vec\sigma=\sigma_x$ with associate Bloch vector $\vec a=(\cos\theta,0,\sin\theta)$ and $\vec b=(1,0,0)$,  the optimal state is the eigenstate of
\begin{equation}
 \vec a\times (\vec a\times \vec b)\cdot\vec\sigma=(1- \cos^2\theta)\sigma_x-\cos\theta\sin\theta\sigma_z
\end{equation}
%
%%%%%%%%%%%%%%%%%%%

\begin{figure}[t!]
\includegraphics[width=8.5cm]{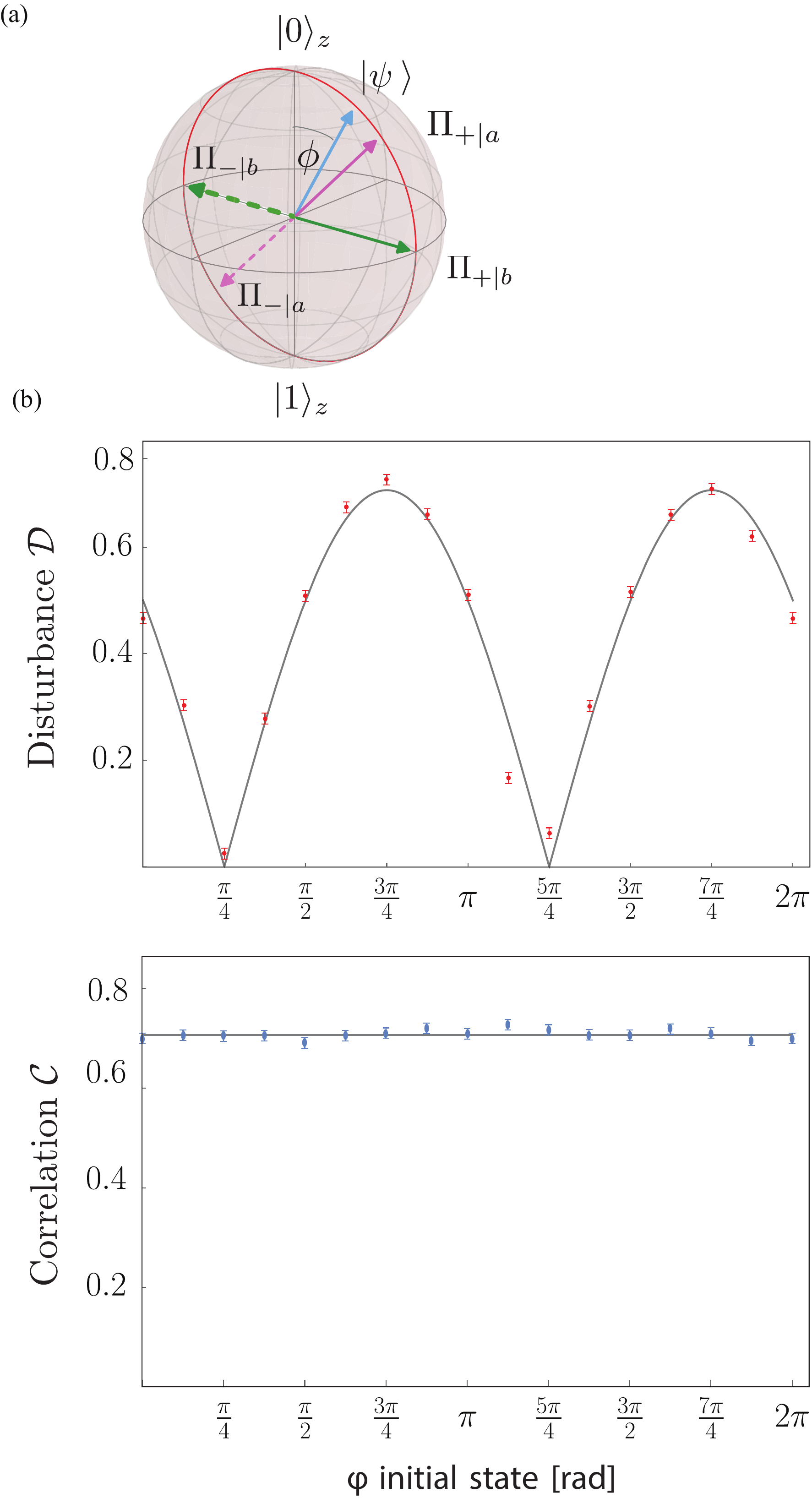}	
\caption{Experimental determination of optimal state $\ket{\psi_{\rm{opt}}(\theta)}$ by varying the polar angle $\phi$ of the initial state $\ket{\psi}$. (a) Bloch sphere description of $\m M_a,\,\m M_b$ and $\ket{\psi}$. (b) Correlation $\m C$ and disturbance $\m D$ versus  polar angle $\phi$ of the initial state $\ket{\psi}$}\label{search} 
\end{figure}

\begin{equation}
 \vec a\times (\vec a\times \vec b)\cdot\vec\sigma=\sin^2\theta\sigma_x-\cos\theta\sin\theta\sigma_z
\end{equation}
with eigenvector,
\begin{equation}\label{eq:optstate}
 \frac{1}{2}(\mathbbm 1\pm\hat n\cdot\vec\sigma) \ \ \ , \ \ \ \hat n=(\sin^2\theta, 0 ,-\cos\theta \sin\theta)/|\vec n|
\end{equation}
with corresponding ket state
\begin{equation}
    \ket{\psi_{\rm opt}(\theta)}=\sin\frac{\theta}{2}\ket{\uparrow_z}+\cos\frac{\theta}{2}\ket{\downarrow_z},
\end{equation}
where $\ket{\uparrow_z}$ and $\ket{\downarrow_z}$ denoted the up and down eigenstates of $\hat\sigma_z$, respectively. Here we present an example of the experimental determination of the optimal state $\ket{\psi_{\rm{opt}}(\theta)}$. Therefore we fix the $\m M_a$ measurement at $\theta=\pi/4$ resulting in $\m M_a=1/\sqrt 2\,\sigma_x+1/\sqrt 2\,\sigma_z$ and vary the polar angle $\phi$ of of the initial state $\ket{\psi}$ between 0 and $2\pi$, which is depicted in Fig.\,\ref{search}\,(a). As seen in Fig.\,\ref{search}\,(b) the maximal disturbance $\m D$ is reached at $\phi=3\pi/4$ and $\phi=7\pi/4$, that is when the initial state is orthogonal to the direction of the $\m M_a$. In addition we see the state-independence of the correlation $\m C$ resulting in a constant value $\m C=1/\sqrt 2$ for all settings of $\phi$ of the initial state $\ket{\psi}$.

\subsection{General (Unsharp) Measurements}

\subsubsection{POVM implementation}

\begin{figure}[h!]
\includegraphics[width=9cm]{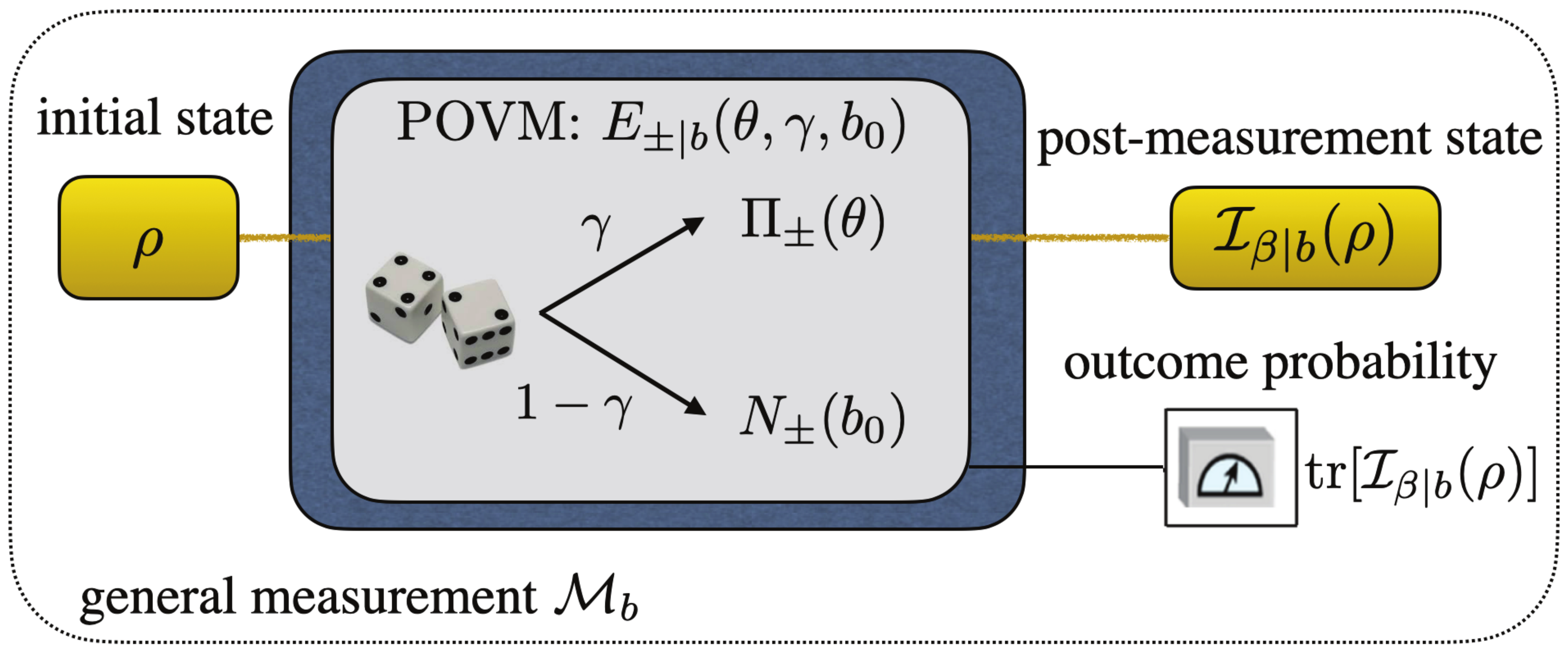}	
\caption{Schematic illustration of general measurement $\m M_a$ described by a set of positive operator-valued
measures (POVMs) $E_\pm (\theta,\gamma,b_0)$. With probability $\gamma$ a projective measurement $\Pi\pm(\theta)$ is carried out and with probability $1-\gamma$ a biased dummy measurement $N_\pm(b_0)$. }\label{POVM} 
\end{figure}

As describes in the main text a general measurements is expressed by 
\begin{equation}\label{eq:POVMbias}
E_{\pm\vert b}(\theta,\gamma,b_0)=(1-\gamma)N_\pm(b_0)+\gamma\Pi_\pm(\theta)
\end{equation}
as a convex combination of projective $\Pi_\pm(\theta)$ and biased dummy measurement $N_\pm(b_0)=\frac{1}{2}(1\pm\frac{b_0}{1-\gamma})\mathbbm 1$, which is fully characterized by three parameters, $\theta$, $\gamma$, and $b_0$, see Fig.\,\ref{POVM}. 
The measurement strength $\gamma=\vert \vec b\vert$ is given by the length of the Bloch vector, furthermore if $b_0=0$ the POVM is unbiased. 

The unbiased POVM we study is given
\begin{equation}\label{eq:POVM}
E_{\pm\vert b}(\theta,\gamma,b_0=0)=(1-\gamma)\frac{{1\!\!1}}{2} +\gamma\underbrace{\frac{1}{2}\Big( {1\!\!1} \pm \overbrace{\cos\theta \sigma_x +\sin\theta \sigma_z}^{\sigma_\theta}\Big)}_{\Pi_{\pm\vert b}(\theta)},
\end{equation}
consisting of a projector $\Pi_{\pm\vert b}(\theta)$ onto the eigenstates of $\sigma_\theta$ with relative weight (probability) $\gamma$, and a contribution of a dummy measurement, as indicated by the unity operator in the POVM definitions from Eq. (\ref{eq:POVM}), with weight $1-\gamma$. In the actual experiment the noisy POVM is realized by controlling the current in the respective coil (if $\m M_b$ is a POVM measurement that would be coil DC-4) with a random generator, switching with the given probabilities $\gamma$ and $1-\gamma$ between the projective measurement and the dummy-measurement. In case of the projective measurement the direction $\theta$ of the projector $\Pi_\pm(\theta)$ is adjusted by the appropriate currents in the coil for the +/- direction. In case of the dummy measurement the current in the coil is switched off, resulting in a random results for the no-measurement.

 A biased POVM in its most compact form is denoted as 
\begin{equation}\label{eq:POVMbias}
E_{\pm\vert b}(\theta,\gamma,b_0)=\frac{1}{2}\big((1\pm b_0){1\!\!1}\pm \gamma(\cos\theta \hat\sigma_x+\sin\theta \hat\sigma_z)\big),
\end{equation}
which for bias $b_0=0$ reproduces Eq. (\ref{eq:POVM}). However, for the actual experimental implementation it is useful to rewrite the POVM in the following form 
\begin{equation}\label{eq:POVMbias}
E_{\pm\vert b}(\theta,\gamma,b_0^\prime)=\gamma\Pi_\pm(\theta)+(1-\gamma)N_\pm(b_0^\prime)
\end{equation}
as a convex combination of projective measurement $\Pi_\pm(\theta)=(\mathbbm 1\pm(\cos\theta \hat\sigma_x+\sin\theta \hat\sigma_z))/2$ and biased dummy (noisy) measurement $N_\pm(b_0^\prime)=(1\pm b_0^\prime)\mathbbm 1/2$, which is fully characterized by three parameters, rotation angle $\theta$ about a specific axes, bias $b_0\prime=b_0/(1-\gamma)$, and the randomization degree $\gamma$. Practically a voltage offset for the control of the current (and consequently the magnetic field) in the DC coils is used to create the biased behavior of the POVM.

First, we study the case where the probe (first) measurement is projective and given by $\m M_{a}=\sigma_x$. In contrast, the target (second) measurement is represented by the unbiased POVM $\m M_{b}(\gamma,\theta)=E_{+\vert b}(\gamma,\theta)-E_{-\vert b}(\gamma,\theta)$. The resulting tight trade off relation $\m C^2+\m D^2=\vert \vec b\vert^2$ can be seen in Fig.\,3 in the main text for different interaction strengths $\gamma=\vert\vec b\vert$.

\subsubsection{Effect of POVM's post-measurement state}

Now we reverse the order of POVM and projective measurement; now the probe (first) measurement is the POVM $\m M_{a}(\gamma)=E_{+\vert a}(\gamma)-E_{-\vert a}(\gamma)$ and the target (second) measurement is projective, given by $\m M_{b}(\theta)=\cos\theta\sigma_x+\sin\theta\sigma_z$ (see Bloch spheres in Fig.\,4 from main text).

\begin{figure}[b!]
\includegraphics[width=9cm]{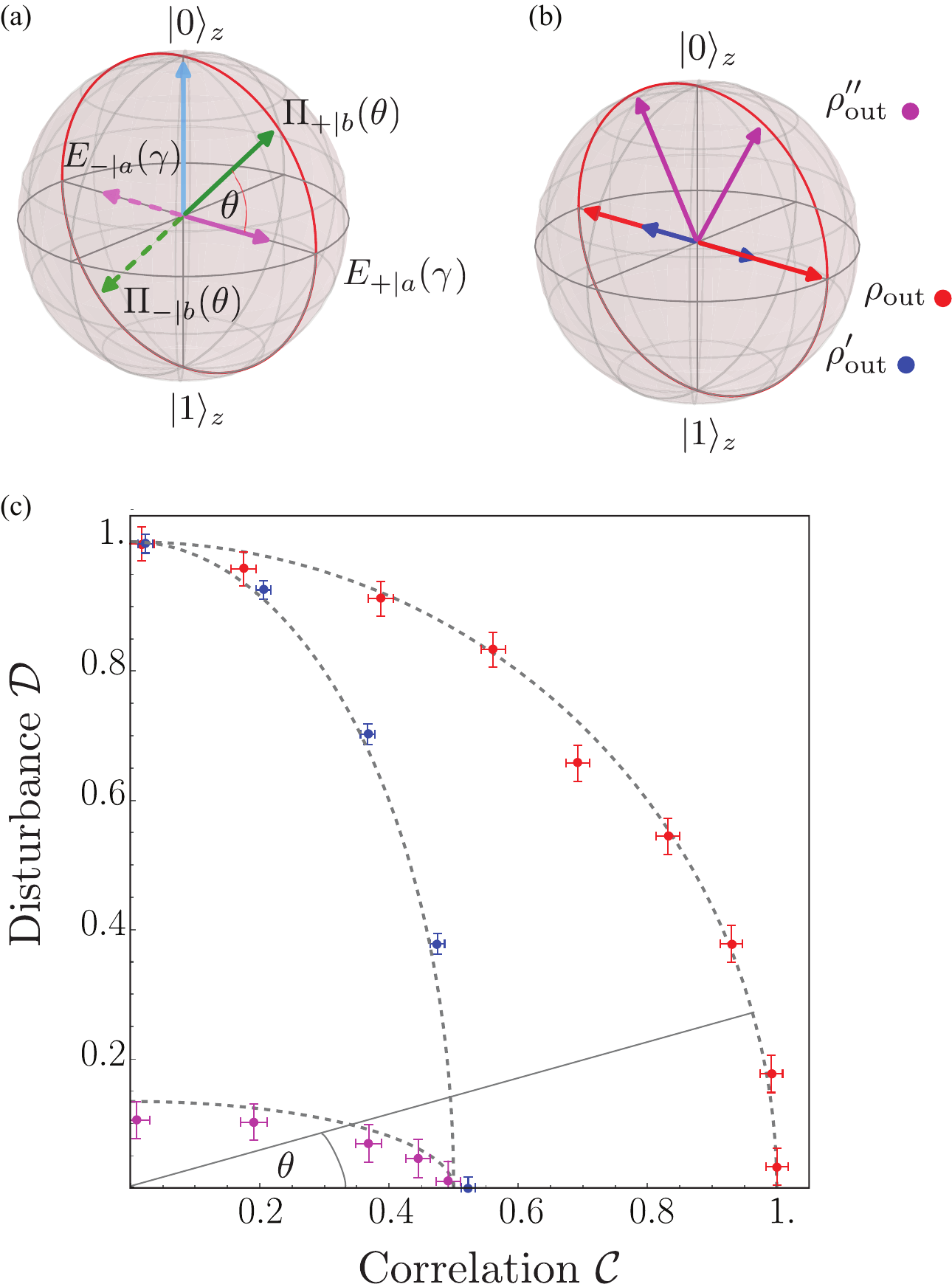}	
\caption{ Effect of POVM's post-measurement state. (a) Bloch sphere representation of measurement settings (b) Bloch sphere representation for post- measurement states pure (red), mixed state (blue) and effect - or L\"uder state-  (purple). (c) Correlation-disturbance tradeoff for the three post-measurement states. }\label{post} 
\end{figure}

The main consequence is that now the {\it{post-measurement}} state of the POVM, denoted as $\rho_{\rm{out}}$ plays a crucial role, representing an additional experimental parameter. Three cases are investigated and the results can be seen in Fig.\,\ref{post}: i) the post-measurement states of the POVM are the eigenstates of the projectors $\rho_{\rm{out}}=\ket{\pm x}$, ii) mixed states with interaction strength $\gamma$ $\rho'_{\rm{out}}=1/2(\mathbbm 1+\gamma\sigma_x)$, and iii) decomposition using the underlying Kraus operators $\rho''_{\rm{out}}=K^\dag_{\pm|a}\rho_{\rm{in}}K_{\pm|a}/{\rm{Tr}}(\hat \Pi_{\pm\vert a}\rho_{\rm{in}})$, with $K^\dag_{\pm|a}K_{\pm|a}=\hat \Pi_{\pm\vert a}$, as required for the so-called L\"uder instrument in the main text.  

%%%%%%%%%%%%%
%%%%%%%%%%%%%%%%%

In the first case $\rho_{\rm{out}}=\ket{\pm x}\bra{\pm x}$, where the post-measurement states $\rho_{\rm{out}}$ equals that from a projective measurement (see Fig.\,\ref{post}\,(a) and (b)), the correlation-disturbance tradeoff relation equals the one of two projective measurements, which is plotted in Fig.\,\ref{post}\,(c) red data points. The reason for this lies in the particular choice of the optimal initial state $\ket{\psi_{\rm{in}}}$, which is perpendicular to the direction of the POVM elements $E_{\pm\vert a}(\theta,\gamma,a_0)$. Hence, the output probabilities are 1/2 for both results - the same as for a projective measurement. Since the post-measurement states $\rho_{\rm{out}}$ equals that from a projective measurement the correlation-disturbance tradeofff relations from projective measurements and POVM with post-measurement states $\rho_{\rm{out}}$ are indistinguishable.

In the second case we study a mixed state 
\begin{equation}
\rho'_{\rm{out}}=1/2(\mathbbm 1+\gamma\sigma_x),
\end{equation}
with interaction strength $\gamma=1/2$ as post-measurement state of the POVM. An interesting behavior is observed: for compatible measurement settings (see Fig.\,\ref{post}\,(a)), that is $\theta=0$ the correlation reproduces the interaction strength $\m C(\theta=0)=0.5=\gamma=\vert \vec a \vert$ (see Fig.\,\ref{post}\,(c) blue data points), as in the case where the measurement order is reversed. However for maximally incompatible orientation of the measurement directions, that is $\theta=\pi/2$ we get disturbance $\m D=1$, which is the same result as for both measurements being projective.

Finally we have the third case, where the corresponding post-measurement state, according to the transformation rule, is described by $\rho\mapsto \m I_{\alpha|a}(\rho)=K_{\alpha|a}\rho K_{\alpha|a}^\dag $ with outcome probability $p_{\alpha|a}=\tr [\m I_{\alpha|a}(\rho)]$ and $E_{\alpha|a}=K_{\alpha|a}^\dag K_{\alpha|a}$ as 
\begin{equation}
\rho''_{\rm{out}}=\frac{K^\dag_{\pm|a}\rho_{\rm{in}}K_{\pm|a}}{{\rm{Tr}}(\hat \Pi_{\pm\vert a}\rho_{\rm{in}})}. 
\end{equation}
Surprisingly the post-measurement state $\rho''$ of the POVM $E_{\pm\vert a}(\gamma)$ is a pure state, due to the particular chioce of the optimal initial state $\ket{\psi_{\rm{opt}}(\theta)}$.  Here an important feature of an unsharp (or general) measurement becomes apparent, namely that the unsharp measurement has less disturbance compared to a projective measurement. For $\theta=\pi/2$ (maximally incompatible)
 the disturbance yields $\m D=0.1$ (Fig.\,\ref{post}\,(c) purple data points), whereas in the case of a projective measurement we have  $\m D=1$ in this case. As seen from Fig.\,4 of the main text the (unbiased) unsharp measurement disturbs the second measurement less compared to a projective measurement.

\subsubsection{Successive unsharp measurements}

\begin{figure}[h!]
\includegraphics[width=8.6cm]{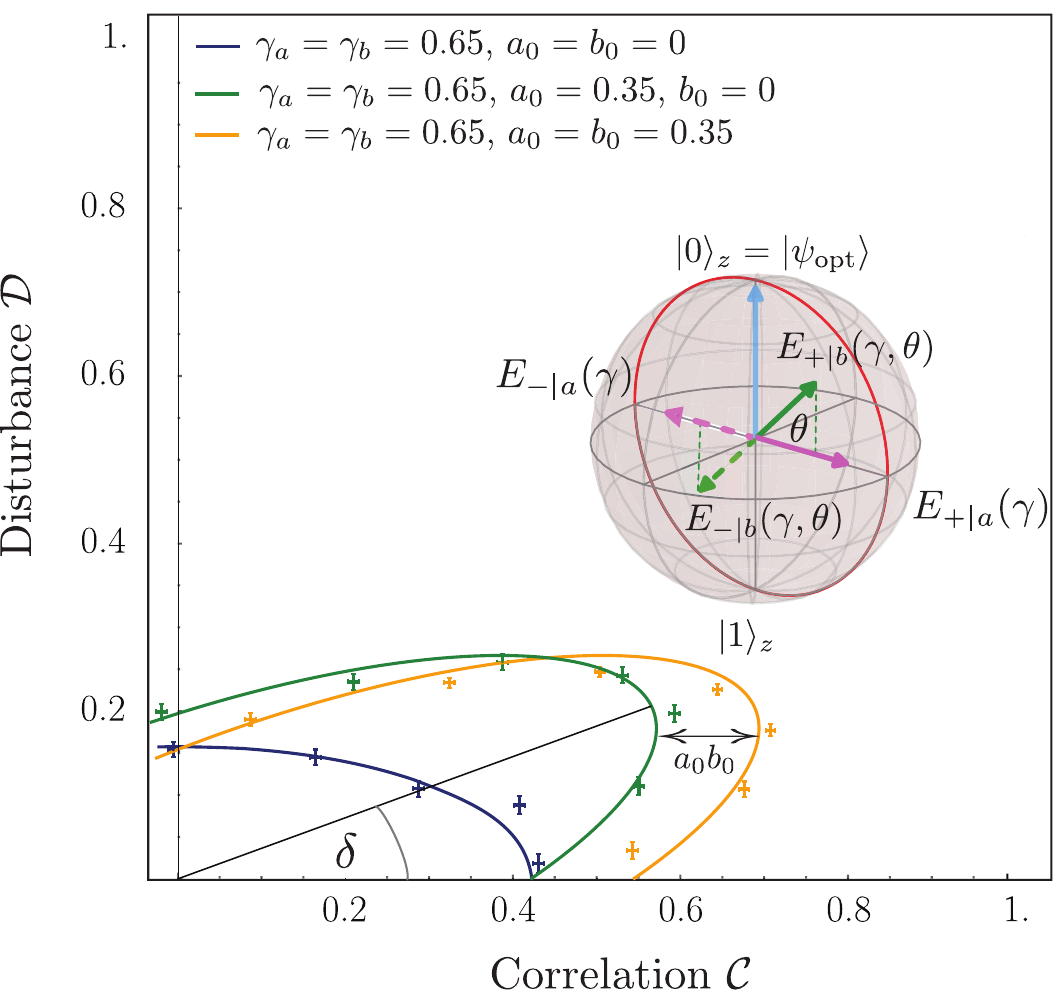}	
\caption{ Effect of POVM's post-measurement bias. blue: no bias, (green) first POVM biased, and (yellow) both POVMs biased. }\label{fig:Succ} 
\end{figure}

Finally, in order to fully demonstrate all properties of the parametric equation from Eq. \ref {eq:ell} we need to perform successive POVM measurements. The probe (first) measurement is given by the (biased/unbiased) POVM $\m M_{a}(\gamma,a_0)=E_{+\vert a}(\gamma,a_0)-E_{-\vert a}(\gamma,a_0)$ and the target (second) measurement is projective, given by represented by the (biased/unbiased) POVM $\m M_{b}(\gamma,\theta,b_0)=E_{+\vert b}(\gamma,\theta,b_0)-E_{-\vert b}(\gamma,\theta,b_0)$, which is depicted in the Bloch sphere representation of Fig.\,\ref{fig:Succ}. The results of the correlation-disturbance tradeoff are plotted in Fig.\,\ref{fig:Succ}. 
%%%%%%%%%%%%%
%%%%%%%%%%%%%%%%%

In blue both POVMs are unbiased ($a_0=b_0=0$), where from $a_0=0$ it follows that $\delta=0$.  As a consequence of $\m M_{a}(\gamma,a_0)$ being unsharp ($\vert \vec a\vert <1$) an ellipse is observed in the correlation-disturbance tradeoff, described by a scaling (squeezing) factor
 \begin{equation}
 s = \frac{1}{2}(a_+ s_+ + a_- s_-), 
 \end{equation}
with
 \begin{equation}
s_\pm = 1 - \sqrt{1 - \frac{|\vec{a}|^2}{a_\pm^2}}.
 \end{equation}
Since $s<\vert \vec a\vert$ we have $\m D_{\rm{max}}=s\,\vert \vec b\vert < \m C_{\rm{max}}=\vert \vec a\vert\,\vert \vec b\vert$; the unsharp measurement disturbs a subsequent measurement less (compared to a projective one cf. Fig.\,3 from the mian text).

In green, where the first measurement is biased ($a_0= 0.35$),  we see a  sheared and tilted ellipse, expressed by shear factor
\begin{equation}
\delta = \frac{1}{2}(u_+ - u_-),
\end{equation}
where
\begin{equation}
u_\pm = \sqrt{a_\pm^2 - |\vec{a}|^2},
\end{equation}
with \(a_\pm = 1 \pm a_0\), caused by the bias of the first POVM given by $a_0$. As a last step, in yellow both POVMs are biased ($a_0=b_0=0.35$), where we observe a horizontal shift of this ellipse by a factor $a_0b_0$, which is the product of the bias of the first and second measurements.

%\bibliography{refs}


\end{document}